\documentclass{vldb}

\usepackage{algorithm}
\usepackage{balance}
\usepackage[noend]{algpseudocode}
\usepackage{caption}
\usepackage{subcaption}
\usepackage{xcolor}
\usepackage{varwidth}
\usepackage{mdwlist}
\usepackage{tabularx}
\usepackage{tikz}
\usetikzlibrary{positioning}
\usepackage{xspace}
\usepackage{url}

\usepackage{marginnote}
\let\oldmarginnote\marginnote
\renewcommand*{\marginnote}[1]{\begingroup \ifodd\value{page}
     \if@firstcolumn\reversemarginpar\fi
   \else
     \if@firstcolumn\else\reversemarginpar\fi
   \fi
   \oldmarginnote{#1}\endgroup }
\makeatletter
\def\@copyrightspace{\relax}
\def\@mkbibcitation{\relax}
\makeatother

\newcommand{\naive}{na\"{\i}ve\xspace}
\newcommand{\naively}{na\"{\i}vely\xspace}

\DeclareMathOperator*{\argmax}{argmax}

\begin{document}

\vldbDOI{https://doi.org/10.14778/xxxxxxx.xxxxxxx}
\vldbVolume{12}
\vldbNumber{xxx}
\vldbYear{2020}
\vldbTitle{Learning a Partitioning Advisor with Deep Reinforcement Learning}
\vldbAuthors{Benjamin Hilprecht et al.}

\title{Learning a Partitioning Advisor\\with Deep Reinforcement Learning}

\numberofauthors{3} 

\author{
\alignauthor
Benjamin Hilprecht\\
       \affaddr{TU Darmstadt}\\
       \affaddr{Germany} 
\alignauthor
Carsten Binnig\\
       \affaddr{TU Darmstadt}\\
       \affaddr{Germany} 
\alignauthor
Uwe R{\"o}hm \\
       \affaddr{University of Sydney}\\
       \affaddr{Australia} 
}

\maketitle

\begin{abstract}
Commercial data analytics products such as Microsoft Azure SQL Data Warehouse or Amazon Redshift provide ready-to-use scale-out database solutions for OLAP-style workloads in the cloud. 
While the provisioning of a database cluster is usually fully automated by cloud providers, customers typically still have to make important design decisions which were traditionally made by the database administrator such as selecting the partitioning schemes.

In this paper we introduce a learned partitioning advisor for analytical OLAP-style workloads based on Deep Reinforcement Learning (DRL). 
The main idea is that a DRL agent learns its decisions based on experience by monitoring the rewards for different workloads and partitioning schemes.
We evaluate our learned partitioning advisor in an experimental evaluation with different databases schemata and workloads of varying complexity. 
In the evaluation, we show that our advisor is not only able to find partitionings that outperform existing approaches for automated partitioning design but that it also can easily adjust to different deployments. 
This is especially important in cloud setups where customers can easily migrate their cluster to a new set of (virtual) machines.
\end{abstract}

\begin{figure*}
    \centering
    \includegraphics[width=0.8\linewidth]{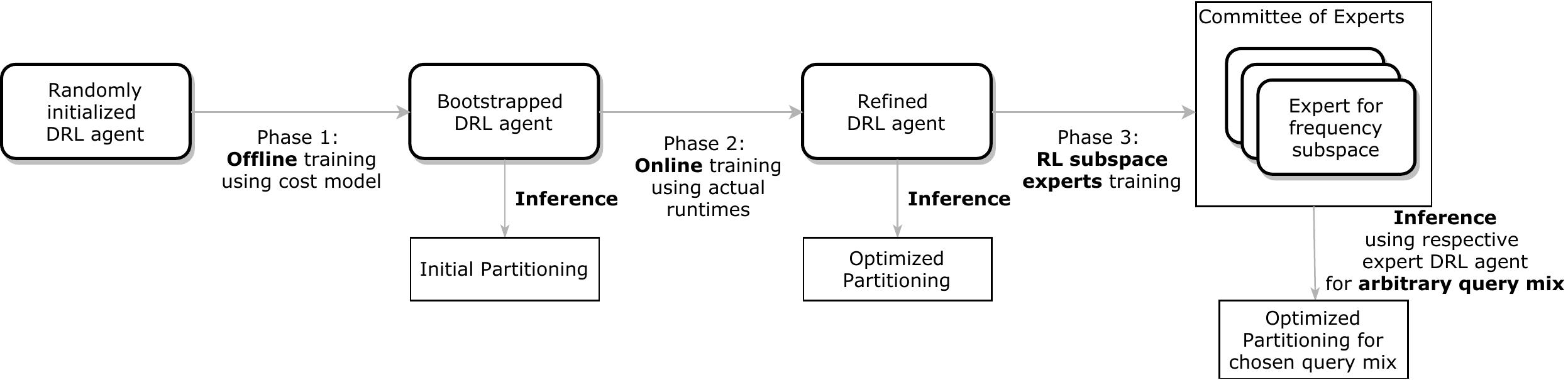}
    \vspace{-1.5ex}
    \caption{Overview of DRL-based approach to Learn a Partitioning Advisor.}
    \vspace{-2.5ex}
    \label{fig:overview}
\end{figure*}

\section{Introduction}
\label{sec:intro}

\paragraph*{Motivation}
Providing data analytics as a service is a growing field in the cloud industry. 
Commercial products such as Microsoft Azure SQL Data Warehouse or Amazon Redshift provide ready-to-use scale-out data warehouse solutions for OLAP-style workloads. 
Using these services, customers can easily define their database schema, upload their data and then query the database using a cluster of machines. 
While the provisioning of the clusters is usually fully automated by cloud providers, customers typically still have to make design decisions which were traditionally made by the database administrator.
For example, in Azure's Data Warehouse but also in Amazon Redshift customers have to decide on the partitioning attribute of a table when creating a new database to split large tables horizontally across multiple machines.
Partitioning the database across nodes can dramatically improve the performance of analytical workloads since data-intensive SQL queries can then be farmed out to multiple machines.

However, partitioning a database in an optimal manner is a non-trivial task and hence different partitioning schemes can significantly impact the overall performance.
For example, analytical queries typically involve multiple joins over potentially large tables. 
If two tables which are co-partitioned on the join attributes have to be joined, the join computation can be executed locally on each node avoiding costly network transfers. 
Deciding for complex schemata with many tables and possible join paths which tables should be co-partitioned is a non-trivial task since this not only depends on the database schemata but also other factors such as table sizes, the query workload (i.e., which joins are actually important and how often tables are joined), or hardware characteristics such as network speed. 

Some techniques have already been proposed to find good data partitionings for analytical workloads over distributed databases \cite{nehme2011automated,DBLP:conf/sigmod/RablJ17}.
However, the existing approaches either rely on heuristics only  \cite{DBLP:conf/sigmod/RablJ17} or solely use cost estimates from the query optimizer \cite{nehme2011automated}, which often do not reflect the real execution costs for more complex queries.

\vspace{-1.5ex}\paragraph*{Contributions}
In this paper, we make the case to use Deep Reinforcement Learning (DRL) for learning an advisor that can suggest good database partitionings.
DRL is particularly interesting compared to other learning setups such as supervised machine learning, since no manually curated training data is required a priori.
Instead, a DRL agent learns its decisions based on experience by trying out different partitioning schemes and monitoring the rewards (i.e., the runtime benefits) for a wide variety of different workloads.

Key to our approach is that we propose an efficient procedure that learns a DRL-based partitioning advisor in two stages. 
First, in an offline phase we use cost estimates for a workload (just as existing approaches) to bootstrap a DRL agent that reflects the basic the trade-offs of using different partitionings for diffent workloads. In a second phase, we then refine the DRL agent by using real execution costs as rewards, which makes the partitioning advisor more accurate than existing approaches that rely only on cost estimates only. 
Once trained, our learned advisor can be queried to obtain a partitioning for a new database or to repartition an already deployed database (e.g., if the workload changes) to a new partitioning scheme that might be better suited for the current mix of queries. 

Another major difference to existing approaches such as \cite{nehme2011automated} is that by making use of DRL, our advisor can not only reflect the trade-offs us using diffent partitioning for different workload mixes in one model but it can also learn the trade-offs for different deployments. 
For example, in Amazon Redshift a customer can migrate its database easily to a new cluster by just deploying a new set of virtual machines that have different hardware characteristics.
Under the new hardware deployment, important factors such as the network bandwidth might change. 
Therefore, a new partitioning scheme is likely to be better suited; e.g., instead of replicating tables, partitioning tables might be more beneficial with increased network speed since network shuffling is now much cheaper and the benefits of partitioning due to a higher degree of parallelism outweigh the costs of data shuffling.
All these trade-offs can be reflected in our learned advisor.

To summarize, we make the following contributions:

\vspace{-1.5ex}
\begin{itemize*}
    \item We first formalize a framework that translates the partitioning problem to a Markov Decision Process making the application of DRL possible. 
    \item We then discuss our two-staged procedure to train the DRL agent using an offline phase that relies on cost estimates to pre-train the model and an online phase that refines the offline-learned advisor by using actual execution costs of a concrete deployment. 
    \item  Moreover, we introduce a committee of DRL agents to improve the adaptivity for flexible workloads. This also allows us to extend the advisor using an incremental training approach if new queries are added to the workload or the database schema changes.
    \item  In our evaluation, we show that our approach can handle a variety of different database schemata and workloads and is able to find non-obvious solutions. Moreover, it adapts its decisions based on workload changes as well as changes in the underlying hardware. 
\end{itemize*}
\vspace{-1.5ex}

\vspace{-1.5ex}\paragraph*{Outline}
The remainder of this paper is organized as follows: 
First, in Section \ref{sec:overview} we provide an overview of our approach to use DRL to learn a partitioning advisor. 
In Section \ref{sec:state_action_definition} we formalize the partitioning problem as a DRL problem before we then introduce our training procedures in Section \ref{sec:training}. 
We then explain how to obtain partitionings at inference time in Section \ref{sec:inference} and present the results of our experimental evaluation in Section \ref{sec:experiments}. 
Finally, we conclude with related work in Section \ref{sec:related_work} and a summary in Section \ref{sec:conclusion}. \section{Overview}
\label{sec:overview}

In this paper, we use an approach based on deep reinforcement learning (DRL) for learning a partitioning advisor. 
The basic idea is that we train a DRL agent that learns the trade-offs of using different partitionings for a given database schema.
In order to allow the DRL agent to generalize to different workloads, we sample a variety of different so called workload mixes (i.e., a set of queries and their frequencies) for learning the advisor.
Once trained, we can use the inference of the DRL agent to derive an ''optimal'' partitioning scheme that aims to minimize the overall runtime for a given workload mix.
In the following, we give an overview of how the training procedure works and how the inference procedure of the learned DRL agent can be used to derive a partitioning. 

Our training procedure of learning the advisor works in multiple phases.
Initially, a DRL agent is trained using an \emph{offline training} phase as shown in Figure \ref{fig:overview} (Phase 1) that uses cost estimates for a given workload as rewards for different partitionings instead of actual query runtimes.
This enables the DRL agent to learn the basic trade-offs between different partitionings without actually executing the workload on a database, which would be too expensive and results in very high runtimes for the training.
Once the training has finished, we can already use the trained DRL agent for inference to obtain the partitioning.

However, when training the DRL agent offline only, the quality of the learned partitionings is bounded by the quality of the cost model similar to existing automated partitioning approaches such as \cite{nehme2011automated}.
The quality of our trained DRL agent is thus further improved by a subsequent \emph{online training} phase as shown in Figure~\ref{fig:overview}  (Phase 2)  that further refines the offline-trained DRL agent. 
For the online training phase, we execute the workload on different partitionings and use the actual runtimes as rewards for refining the decisions of the DRL agent. 
As we show in our experimental evaluation in Section~\ref{sec:experiments}, this helps the DRL agent to find better partitionings than an offline-trained agent and allows us to adopt the agent to the characteristics of a given deployment (e.g., different network speeds).

Once the training finished (offline and online), we can use the \emph{inference procedure} of the DRL agent to decide which partitioning should be used.
The main idea is that based on a given workload mix (i.e., queries and their frequencies), the DRL agent suggests a partitioning that aims to minimize the runtime for the workload mix.
Once trained, the learned DRL agent can then be used to suggest an initial partitioning for a new database and a mix of queries or to adapt the partitioning of an already deployed database if the workload mix changes.
However, while we are capable of adapting the partitioning based on changes in the workload, it is beyond the scope of this paper to monitor or predict the workload mix and decide whether the overall runtime savings are potentially worth the additional efforts used for repartitioning the database.

An important extension of our approach is that we train not only one but several DRL agents which form a committee of experts. Instead of training a single agent which learns the trade-offs for all possible workload mixes, one DRL agent is solely the expert for a subspace of the workloads (i.e., particular mixes of queries) as shown in Figure~\ref{fig:overview} (Phase 3).
Given a new workload mix, we can use the inference of the dedicated expert for the workload subspace to obtain a partitioning which further improves the quality of our advisor as shown in our experiments in Section~\ref{sec:experiments}.

A last important aspect is that we are able to efficiently adapt the learned advisor by incremental training if completely new queries are added to the workload or the database schema changes.
However, for supporting a completely new database schema (as well as a workload for that schema) a new set of DRL agents must be trained.
To better support this case in future, we plan to investigate transfer learning approaches to reuse already trained DRL agents to speed-up the training for new database schemas.
Using transfer learning is in particular interesting for cloud providers that see a large number of different schemas and thus can select from a wide variety of learned agents. \section{Partitioning as a DRL Problem}
\label{sec:state_action_definition}

In the following, we discuss the required background on Deep Reinforcement Learning (DRL) before we show how the partitioning problem can be formulated as a DRL problem.
At the end, we discuss the benefits and limitations of using DRL compared to other alternative approaches for automatic partitioning.

\subsection{Background}

In Reinforcement Learning (RL), an agent interacts with an environment by choosing actions. 
Specifically, at each discrete time step $t$, the agent observes a state $s_t$. By choosing an action $a\in A,$ it transitions to a new state $s_{t+1}$ and obtains a reward $r_t$. 
Mathematically, this can be modeled as Markov decision process. The way the agent picks the actions depending on the state is called policy $\pi.$ 
The goal of the agent is to maximize the rewards over time. 
However, the greedy policy, i.e. selecting the action with the highest immediate reward, might not be the best strategy. Instead, the agent might select an action that enables higher rewards in the future. 
Consequently, when selecting actions the agent should always keep the long-term rewards in mind \cite{sutton2018reinforcement}. 

One approach to solving this problem is Q-learning. With the Q-function, the expected discounted future rewards are approximated as follows, if we pick action $a$ at state $s$:

\begin{equation*}
Q(s,a) = \mathbb{E}\left(\sum_{t=0}^{\infty}r_t(s_t,a_t) \gamma^t | s_0=s, a_0=a\right)
\end{equation*}

The rewards are discounted with a factor $\gamma<1$ to account for a higher degree of uncertainty for future states. 
The $Q$-function is learned during training. 
Note that if the approximation is good enough we can choose an optimal action for a state $s$ as $\argmax_{a \in A}Q(s,a)$.

During training one also has to select random actions so that there is a trade-off between exploration and exploitation. 
Usually, exploration is realized by picking a random action with probability $\epsilon.$ 
This probability is decreased over time \cite{sutton2018reinforcement} by multiplication with a factor called \emph{epsilon decay}.

There are different ways of realizing the $Q$-function. For Deep Q-learning \cite{mnih2015human} (or Deep Reinforcement Learning), a neural network $Q_{\theta}(s,a)$ with weights $\theta$ is used for the approximation. 
Having observed a state $s_t$ and an action $a_t$, the corresponding immediate reward $r_t$ and the future state $s_{t+1}$ the network is updated via Stochastic Gradient Descent (SGD) and the squared error loss:

\begin{equation*}
\label{eqn:DQN_learning}
\left(r_t + \gamma \argmax_{a\in A}Q(s_{t+1},a_t)-Q(s_t,a_t)\right)^2
\end{equation*}

The intuition is that the expected discounted future rewards when selecting action $a_t$ in step $t$ should be the immediate reward $r_t$ together with the maximum expected discounted future rewards when selecting the best action $a$ in the next step $t+1$ discounted by $\gamma,$ i.e. $\argmax_{a\in A}Q(s_{t+1},a_t)$.

\begin{figure*}
    \centering
    \begin{minipage}{.45\textwidth}
        \includegraphics[width=0.8\linewidth]{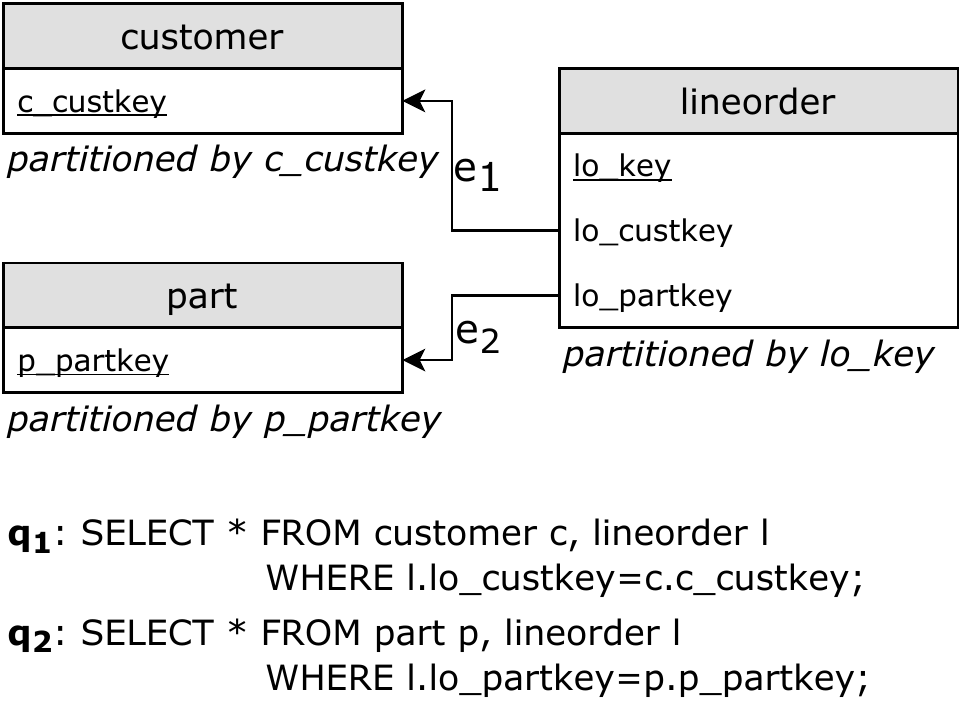}
        \subcaption{Schema and Workload}
        \label{fig:example_state_ssb:schema}
    \end{minipage}
    \hfill
    \begin{minipage}{.45\textwidth}
        \begin{scriptsize}
        \texttt{lineorder(lo\_key,lo\_custkey,lo\_partkey)}\\
        partitioned by \texttt{lo\_custkey}\\
        $s(\mathit{lineorder})=\begin{pmatrix}r_1,a_{11},a_{12},a_{13}\end{pmatrix}=\begin{pmatrix}0,0,1,0\end{pmatrix}$\\
        \newline
        \texttt{customer(c\_custkey)}\\
        partitioned by \texttt{c\_custkey}\\
        $s(\mathit{customer})=\begin{pmatrix}r_2,a_{21}\end{pmatrix}=\begin{pmatrix}0,1\end{pmatrix}$\\
        \newline
        \texttt{part(p\_partkey)}\\
        partitioned by \texttt{p\_partkey}\\
        $s(\mathit{part})=\begin{pmatrix}r_3,a_{31}\end{pmatrix}=\begin{pmatrix}0,1\end{pmatrix}$\\
        \newline
        Two queries occurring equally frequently\\
        $s(Q)=\begin{pmatrix}f_1,f_2\end{pmatrix}=\begin{pmatrix}1,1\end{pmatrix}$\\
        \newline
        Edge $e_1$ for \texttt{lo\_custkey$\rightarrow{}$c\_custkey}: active \\
        Edge $e_2$ for \texttt{lo\_partkey$\rightarrow{}$c\_partkey}: inactive \\
        $s(E)=\begin{pmatrix}e_1,e_2\end{pmatrix}=\begin{pmatrix}1,0\end{pmatrix}$\\
         \vspace{-1.5ex}
         \subcaption{State Representation}
        \label{fig:example_state_ssb:state}
        \end{scriptsize}
    \end{minipage}
    \vspace{-1.5ex}
    \caption{State Representation of Simplified SSB Schema and Workload.}
    \label{fig:example_state_ssb}
    \vspace{-1.5ex}
\end{figure*}

\subsection{Problem Formulation}

A partitioning problem can be more formally described as follows:
Given a set of tables $T$ and queries $Q$, for every table $T_i\in T$ with attributes $a_{i1},a_{i2},\dots,a_{in}$ we have to decide whether to replicate or to partition the table. 
For simplicity, in the following, we assume that we use only one partitioning scheme (e.g., hash-partitioning) that horizontally splits a table into a fixed number of shards (which is equal to the number of nodes in the database cluster).
Thus, when two tables use the same partitioning attribute, we say that they are co-partitioned on that attribute and equi-joins on that attribute do not involve any data shuffling.
For replication, we also make a simplifying assumption: if the table is selected to be replicated, it is replicated to all nodes in the cluster.
To summarize, for each table, we have to decide whether that table is partitioned or replicated. If it is partitioned we additionally need to decide which attribute $a_i$ (or set of attributes) is being used for horizontal partitioning.

The main intuition to model the partitioning as a DRL problem is to model the database and the workload as state and possible changes in the partitioning as actions.
In the following, we thus first show how we model states and actions and how we select an appropriate reward.
We explain our representation of states and actions using a simplified SSB schema as shown in Figure \ref{fig:example_state_ssb}.

\vspace{-1.5ex}\paragraph*{State} The most important part of the state is to model the current database schema and the selected partitioning; i.e., for every table $T_i$ we have to encode whether it is currently partitioned or replicated. If it is partitioned, we additionally have to encode which attribute(s) are used for partitioning. 
Hence, for each table, we can encode its state as binary vector using one-hot encoding:

\begin{equation*}
\label{eqn:state_per_table}
s(T_i)=\begin{pmatrix}
r_i, a_{i1}, a_{i2}, \dots, a_{in}
\end{pmatrix}
\end{equation*}

where $r_i$ encodes whether a table is replicated or not and the remaining bits indicate whether an attribute is used for partitioning.

Moreover, we also need to model the workload as part of the state since for the same database schema, different workloads result in different partitioning strategies that should be selected.
For now, we assume that every possible query $q_i$ in a workload of queries $Q$ is known in advance which is not uncommon in OLAP workloads.
In the next section, we discuss how the workload can be extended with new queries.
To encode a workload, we again represent it as a vector where an entry encodes the current frequency $f_i$ of a query $q_i:$

\begin{equation*}
\label{eqn:freq_vector}
s(Q)=\begin{pmatrix}
f_1, \dots, f_{m}
\end{pmatrix}.
\end{equation*}

That way, the input state can represent different query mixes. For example, if a query occurs twice as often as another query its frequency is twice as high.
For representing the frequency of a query, we normalize $f_i,$ i.e. it is divided by the highest frequency in a workload.
Hence we obtain a vector where the entries for $f_i$ are in the interval $[0,1]$. 

To limit the search space and prevent the exploration of sub-optimal partitionings, we further extend the state representation pushing the agent to explore only partitionings where tables are co-partitioned. 
Therefore, we introduced the concept of edges. 
Every pair of join attributes $a_{ir}$ and $a_{js}$ of the corresponding tables $T_i$ and $T_j$ induces an edge $(T_i,a_{ir},T_j,a_{js})$ which can either be active or inactive. 
If it is active, the two corresponding tables are guaranteed to be co-partitioned if we join them on this relationship, i.e. $T_i$ is hash-partitioned by $a_{ir}$ and $T_j$ is hash-partitioned by $a_{js}$. 
For example, in Figure \ref{fig:example_state_ssb} the edge $e_1$ is active and thus the \texttt{lineorder} and \texttt{customer} are partitioned by the attributes \texttt{lo\_cust\-key} and \texttt{c\_custkey}, respectively.

The set of possible edges $E$ can easily be extracted from the given workload (i.e, all possible join paths). 
As every edge can either be active or inactive, the edge states can also be represented as a one-hot encoded vector

\begin{equation*}
s(E)=\begin{pmatrix}
e_1, \dots, e_{m'}
\end{pmatrix}.
\end{equation*}

\vspace{-1.5ex}\paragraph*{Actions} A small state space is essential to apply $Q$-learning because we have to compute the $Q$-values for all possible actions to decide which action to execute in a state.
Hence, we designed all actions to change only one aspect of the partitioning encoded in the state at a time.

Consider, for simplicity, a single table $T_i$: Possible actions are to replicate the table or to partition the table by one of its attributes $a_{ir}$. 
These actions can only be selected if the table is currently in a different physical design, e.g. to replicate table $T_i$ is only a valid action if the table is not yet replicated. 
Moreover, we restrict the actions to those which do not conflict with the current set of active edges. 
For example, replicating table $T_i$ is not a valid action if the edge $(T_i,a_{ir},T_j,a_{js})$ is active because it guarantees that $T_i$ is co-partitioned with $T_j$ using $a_{ir}$ and $a_{js}$ as partitioning attributes.

The second class of actions considers activating/de-activating edges as short-cuts to change the partitioning.
Intuitively, activating an edge co-partitions two tables while the de-activation of edges allows follow-up actions to choose a new partitioning attribute for that table.
It is important that the set of edges to be activated is conflict-free. For this, we solely have to check that in the set of edges there are no two edges which require a table $T_i$ to be partitioned by different attributes $a_{ir}$ and $a_{ir'}$.
For example, edge $e_2$ cannot be activated in Figure \ref{fig:example_state_ssb} because $e_1$ is already active.

\vspace{-1.5ex}\paragraph*{Rewards} The overall goal of the learned advisor is to find a partitioning that minimizes the runtime for the workload mix (queries and their frequencies) modeled as part of the input state. 
This objective has to be minimized by the DRL agent and can be used as a reward. 
We can either directly use the actual runtimes $c_r(P,q_i)$ of the queries $q_i$ given a partitioning $P$ or alternatively use a cost model $c_m(P,q_i)$ (e.g., of a query optimizer) approximating the true runtime. 
In fact, we use both costs: estimates of the cost model are used for the offline training and actual costs for the online training. 
For simplicity, but without loss of generality, in the following we just assume one cost function $c(P,q_i)$ that reflects the runtime of the query $q_i$ over the partitioning $P$. 

Since the DRL agents seeks to maximize the reward, we use negative costs in the reward definition. Instead of simply defining the reward to be $-\sum_{j=1}^{m} f_i c(P,q_j)$, we additionally normalize the reward with respect to the runtime of some reference partitioning $P_{0}$ (e.g., $P_{0}$ can be a \naive{} partitioning where all tables are partitioned by their primary key):

\begin{equation*}
\label{eqn:reward_dynamic}
r=-\frac{\sum_{j=1}^{m} f_j c(P,q_j)}{\sum_{j=1}^{m} f_j c(P_{0},q_j)}.
\end{equation*}

That way, we can use estimated costs from a cost model to pre-train the DRL agent and then in later phases change the reward to use actual execution costs.

\subsection{Discussion and Limitations}

It turns out that formalizing the partitioning problem as DRL problem has several advantages. 
First, training an DRL agent mimics the way how a database administrator approaches the problem. 
Both the administrator and the DRL agent first try different partitioning approaches to explore the trade-offs of the design space. 
That way, a learned partitioning advisor can better cope with the complexity of the partitioning problem than existing approaches and can take effects into account that are particular to a certain setup (e.g., network speed).

As discussed before, for training an RL agent we assume that the workload (i.e., the set of queries but not the frequencies) is known which is common for many OLAP applications that run a fixed set of reports. 
However, different workload mixes of the given set of queries are supported by using different frequency vectors  as input (i.e., if some queries occur more often than others) without the need to retrain the agent.
Instead, as mentioned before, the RL agent is already trained with a variety of different workload mixes and the RL agent can thus be used to predict the partitoning also when the workload mix changes.
Moreover, when adding a completely new query to the state space, we do not have to train a new DRL agent from scratch. 
Instead, we can simply extend the input space of the already trained DRL agent and retrain the learned advisor by sampling new frequency vectors that include the new queries. 
As we show in our experiment in Section \ref{sec:experiments}, the time required for retraining the DRL agent for completely new queries is only a small fraction of the original training time.

Beyond adding new queries, the schema might change or new data might be inserted into the database which changes the costs for a workload and hence the partitioning strategy. 
For new tables (or schema changes in general) the same observation as before holds; i.e., we only need to retrain the current agent which takes much less time than training an agent from scratch.
Furthermore, if the new data is inserted into the database that follows the same characteristics, e.g. the ratios of the table sizes do not change significantly we can continue to use the DRL agent trained on the initial database. 
Only if the data distribution changes significantly, we need to retrain the agent again in an incremental manner to reflect the changed execution costs. \section{Training Procedure}
\label{sec:training}

In the following, we discuss the details of the offline and the online phase of our DRL-based training procedure.
At the end of this section, we further discuss optimizations of our approach that allow us to provide a higher accuracy for changing workloads (i.e., if the frequency of queries change) and to incrementally add new unseen queries (and tables) to the workload mix.

\subsection{Phase 1: Offline Training}

For training a partitioning advisor, the DRL agent interacts with the state reflecting the current partitioning by selecting different actions and observing rewards as described before in Section \ref{sec:state_action_definition}. 
The actions in our model change the current partitioning and the rewards represent the total cost to run a given mix of queries on that partitioning. 
For the offline training, the runtimes of different partitionings $P$ are only approximated with a cost model $c_m(P,q_i)$. 
The advantage over actual runtimes is that we do not have to execute queries on an actual database.
However, the disadvantage is that the quality of the  partitioning found by the DRL agent is bounded by the accuracy of the cost model. If, for example, the optimal partitioning $P^\ast,$ i.e. one that minimizes the runtime on the cluster $c_r(P^\ast,q_i)$, has higher costs $c_m(P^\ast,q_i)$ than some other partitioning $c_m(P',q_i)$ the model would favor the sub-optimal partitioning $P'$. 
We overcome this inaccuracy by further training the DRL agent in an online manner as discussed in the next section.

It is a natural approach to use the internal cost model of the database as cost model for the offline phase. 
Obviously, this only works if it accurately reflects the effect of different partitionings. 
Hence, we can only use the internal cost model of the database if we can obtain cost estimates without actually partitioning the data. 
In this paper, we use a cost model $c_m(P,q_i)$ as implemented by query optimizers of distributed databases. By factoring in computation and network transfer cost it allows us to estimate the overall execution costs of a given query $q_i$ for a partitioning strategy.
In our experiments, we show that based on this simple cost model, we can already train an DRL agent that is able to suggest good partitionings.

\begin{algorithm}[!t]
 \scriptsize
\caption{Offline Training}\label{alg:offline_training}
\begin{algorithmic}[1]

\State Randomly initialize Q-network $Q_{\theta}$
\State Randomly initialize target network $Q_{\theta'}$
\For{e in $0,1,\dots,e_{\max}$} \Comment{Episodes}
    \State Reset to state $s_0$
    \For{t in $0,1,\dots,t_{\max}$} \Comment{Steps in Episode}
        \State Choose $a_t=\argmax_{a}Q_{\theta}(s_{t+1},a)$ with \par 
        \hskip\algorithmicindent probability $1-\varepsilon,$ otherwise random action 
        \State Execute action $a_t$ (i.e., simulate what the next \par 
        \hskip\algorithmicindent state $s_{t+1}$ and partitioning $P_{t+1}$ would be)
        \State Compute reward with cost model $c_m$: \par 
        \hskip\algorithmicindent $r_{t}=\frac{\sum_{j=1}^{m} f_j c_m(P_{t+1},q_j)}{\sum_{j=1}^{m} f_j c_m(P_{0},q_j)}$
        \State Store transition $(s_t,a_t,r_t,s_{t+1})$ in B
        \State Sample minibatch $(s_i,a_i,r_i,s_{i+1})$ from B
        \State Train Q-network with SGD and loss \par 
        $\sum_{i=1}^{b} (r_i + \gamma \argmax_{a\in A}Q_{\theta'}(s_{i+1},a)-Q_{\theta}(s_i,a_i))^2$
    \EndFor
    \State Decrease $\varepsilon$
    \State Update weights of target model: $\theta'=(1-\tau)\theta'+\tau \theta$
\EndFor
\end{algorithmic}
\end{algorithm}

Using our cost model as well as the state/action representation introduced before, we can train the DRL agent as described in Algorithm~\ref{alg:offline_training}. 
The training is divided into sequences of states and actions of length $t_{\max}$ called episodes. The selected actions change the partitioning used for the cost estimation $c_m(P,q_i)$. At the end of the episode the DRL agent returns to the state $s_0.$ 
This state $s_0$ has to be some fixed initial partitioning, e.g., partition every table by its primary key. 
To guarantee that the DRL agent can find the optimal partitioning we have to make sure that it can reach any other partitioning within $t_{\max}$ steps starting from the reference partitioning $s_0$. Since for every table we need just one action to partition it by any attribute or replicate it, any state can be reached within at most $|T|$ actions (where $|T|$ denotes the number of tables in the schema). However, as $t_{\max}$ influences the training time it is also a hyperparameter that needs to be optimally set as any other hyperparameter. 
We will later discuss how we selected $t_{\max}$ for schemas with different complexity in our experiments.

\subsection{Phase 2: Online Training}
\label{sec:online_training}

In contrast to offline training, the idea of online training is to deploy the partitionings $P_i$ on a database cluster and measure the true runtimes $c_r(P_i,q_i)$ to compute the reward. 
However, the \naive{} approach is way too expensive to be used in practice. Imagine for example that we need $1200$ episodes for training each having $t_{\max}=100$ steps. Assume we have just a few queries and a small schema such that the total workload takes around 20 minutes and the repartitioning takes another 20 minutes on average. If we simply executed every action the agent proposes, i.e. we repartition the tables and measure the workload runtimes on the cluster, we would end up with a runtime of $(20\mathit{mins}+20\mathit{mins})*1200*100\approx 9 \mathit{years}.$ 

Therefore, online training is intended to (only) serve as refinement in addition to offline training. This has no effect if we use the same degree of exploration, i.e. if we choose random actions with the same probabilities $1-\varepsilon$. 
Note that $\varepsilon$ is multiplied with a certain factor called \emph{epsilon decay} after every episode to decrease it over time. 
For online training, we start with the $\varepsilon$ value that we would reach after 600 episodes (i.e. half of the usual amount of episodes) in the offline phase. This already significantly reduces the training costs as we will show in our experiments. 
However, this does not suffice to effectively reduce the time of the online phase in practice. 
We therefore use further optimizations as discussed next. 

\vspace{-1.5ex}\paragraph*{Sampling} Instead of using all tuples of a database, we just use a sample for every table. This speeds up both the runtime of the queries and the time needed to repartition or replicate any table. In addition, we found it useful not to use the runtimes of a query $c_r(P,q_i)$ directly but to multiply this with a certain factor for every query. The intuition is that some queries scale better than others on the full dataset. Hence, runtime improvements of queries that scale better and thus also run fast on the full dataset should weigh lower than improvements of queries which are very slow on the full dataset. To this end, we measure the runtimes of each query~$q_i$ for the partitioning $P_{\mathit{offline}}$ found in the offline phase once for the full dataset $c_{\mathit{full}}(P_{\mathit{offline}},q_i)$ and once for the sample $c_{\mathit{sample}}(P_{\mathit{offline}},q_i).$ Afterwards, we scale the costs for each query $q_i$ with the corresponding factor $s_i=\frac{c_{\mathit{full}}(P_{\mathit{offline}},q_i)}{c_{\mathit{sample}}(P_{\mathit{offline}},q_i)}.$

One question is how many tuples have to be sampled per table, i.e. how the sampling rate is chosen. 
Higher sampling rates result in a longer runtime of the online phase since both the query runtimes as well the repartitioning times will increase. In contrast, smaller sampling rates might lead to suboptimal partitionings. This can happen if partitionings $P'$ have shorter weighted runtimes $s_i c_{\mathit{sample}}(P',q_i)$ on the sample than the optimal $P^\ast$ for the full dataset. 
We can account for these cases by selecting several partitionings $P_1,\dots,P_n$ and measure their runtime both for the sample and for the full dataset. 
If partitionings with shorter weighted runtimes on the sample also lead to shorter runtimes on the full dataset size the sampling rate is sufficient. If not, the sample size has to be increased.

As a simple heuristic one can empirically determine a threshold below which table sizes should not fall after sampling. This guarantees that tables have a certain minimum size. If this threshold is large enough, optimal partitionings on the samples will also be optimal on the full dataset with high probability. 
A cloud provider could empirically determine this threshold for every database and hardware setup.

\vspace{-1.5ex}\paragraph*{Query Runtime Caching} If the DRL agent visits two states $s_i$ and $s_j$ during training which have the same corresponding partitioning $P$, then the rewards $r_i$ and $r_j$ must consequently be identical. Hence, we can cache query runtimes to faster compute recurring reward values. Additionally, if the partitionings of the states $s_i$ and $s_j$ differ only for a certain set of tables $\{T_{i1},T_{i2},\dots,T_{in}\}$ we only have to measure the runtimes of queries $q_i$ that contain at least one of these tables. In particular, the runtime of every query $q_i$ containing the tables $\{T_{i1},T_{i2},\dots,T_{in}\}$ depends only on the states of these tables, i.e. $s(T_{i1}),s(T_{i2}),\dots,s(T_{in})$. Hence, for every query we can maintain a table containing the different state combinations $s(T_{i1}),s(T_{i2}),\dots,s(T_{in})$ and the runtime of the query on the sample dataset. In summary, when visiting a new state we examine the state of every table $s(T_i)$; i.e. whether it is replicated or hash-partitioned by a certain attribute, and run only the queries $q_i$ for which we do not have a runtime entry for the state combination of relevant tables $s(T_{i1}),s(T_{i2}),\dots,s(T_{in})$.

\begin{algorithm}[!t]
 \scriptsize
\caption{Online Training}\label{alg:online_training}
\begin{algorithmic}[1]

\State Sample from full dataset and deploy tables with initial partitioning on cluster
\State Initialize $P_{\mathit{actual}}$
\State Randomly initialize Q-network $Q_{\theta}$
\State Randomly initialize target network $Q_{\theta'}$
\State Initialize Query Runtime Cache $C=\{\}$
\For{e in $0,1,\dots,e_{\max}$} \Comment{Episodes}
    \State Reset to state $s_0$
    \For{t in $0,1,\dots,t_{\max}$} \Comment{Steps in Episode}
        \State Choose $a_t=\argmax_{a}Q_{\theta}(s_{t+1},a)$ with \par 
        \hskip\algorithmicindent probability $1-\varepsilon,$ otherwise random action 
        \State Execute action $a_t,$ i.e. simulate what the next \par 
        \hskip\algorithmicindent state $s_{t+1}$ and Partitioning $P_{t+1}$ would be
        \State Determine set of queries $Q_t$ which have to be run \par 
        \hskip\algorithmicindent on $P_{t+1}$ because of missing values in C
        \State // \textit{Lazy Repartitioning}
        \State Determine set of tables $T_{Qt}$ contained in $Q_t$
        \For{$T_i$ in $T_{Qt}$}
            \If{$T_i$ is different in $P_{\mathit{actual}}$ and $P_{t+1}$}
                \State Repartition $T_i$ according to $P_{t+1}$
                \State Update $P_{\mathit{actual}}$
            \EndIf
        \EndFor
        \State // \textit{Fill Query Runtime Cache}
        \For{$q_i$ in $Q_t$}
            \State Measure $t=c_{\mathit{sample}}(P_{t+1},q_i)$ by running $q_i$ \label{alg:line:timeout}
            \State Determine states of relevant tables for $q_i$: \par 
            \hskip\algorithmicindent\hskip\algorithmicindent $\begin{pmatrix}s(T_{i1}),\dots,s(T_{in})\end{pmatrix}$
            \State Store $(i,\begin{pmatrix}s(T_{i1}),\dots,s(T_{in})\end{pmatrix},t)$ in C
        \EndFor
        \State // \textit{Exploit Query Runtime Cache}
        \State Lookup costs $c_{\mathit{sample}}(P_{t+1},q_i)$ in C
        \State Compute reward:  $r_{t}=\frac{\sum_{j=1}^{m} f_j s_j c_{\mathit{sample}}(P_{t+1},q_j)}{\sum_{j=1}^{m} f_j s_j c_{\mathit{sample}}(P_{0},q_j)}$
        \State Store transition $(s_t,a_t,r_t,s_{t+1})$ in B
        \State Sample minibatch $(s_i,a_i,r_i,s_{i+1})$ from B
        \State Train Q-network with SGD and loss \par 
        $\sum_{i=1}^{b} (r_i + \gamma \argmax_{a\in A}Q_{\theta'}(s_{i+1},a_i)-Q_{\theta}(s_i,a_i))^2$
    \EndFor
    \State Decrease $\varepsilon$
    \State Update weights of target model: $\theta'=(1-\tau)\theta'+\tau \theta$
\EndFor
\end{algorithmic}
\end{algorithm}

\vspace{-1.5ex}\paragraph*{Lazy Repartitioning} Imagine a simple example where the agent first chooses the action to replicate the table $T_0$. Due to Query Runtime Caching it is possible that we do not have to measure any runtimes for queries $q_i$ where $T_0$ is involved because we have already stored the runtimes for all these queries and the current state. Hence, we can compute the reward without having to execute any query. Further assume that the agent decides to partition the table $T_0$ by some other attribute afterwards. In this case it was unnecessary to replicate $T_0$ on the actual cluster in the first place. In general, we only have to repartition a table, if runtimes for queries where this table is involved have to be measured.

The approach of lazy repartitioning is to keep track of the partitioning deployed on the database $P_{\mathit{actual}}$ and the partitioning $P_{t}$ of the state $s_t$ the agent is currently at. Every time the agent chooses an action and we reach a new state we first check which queries $\{q_{j1},\dots,q_{jn}\}$ have to be executed on the database. Especially in later phases of training this will be significantly fewer queries than the full set $Q$ since many runtimes will be in the Query Runtime Cache. For this set we determine the set of tables $\{T_{i1},\dots,T_{im}\}$ which are contained in these queries. For every table we check, if the actual partitioning on the cluster matches with the partitioning $P_{t}$ of the current state $s_t$. Only if these do not match, we actually repartition the table.

\vspace{-1.5ex}\paragraph*{Timeouts} The idea of this optimization is that a partitioning where a single query exceeds a certain time limit cannot be optimal. Hence, we can safely abort the query execution and move on with training. The reward for a partitioning $P$ for online training is defined to be

\begin{equation*}
r=-\frac{\sum_{j=1}^{m} f_j s_j c_{\mathit{sample}}(P,q_j)}{\sum_{j=1}^{m} f_j s_j c_{\mathit{sample}}(P_{0},q_j)}=-\frac{\sum_{j=1}^{m} f_j s_j c_{\mathit{sample}}(P,q_j)}{C}.
\end{equation*}

We can also compute the reward $r_{\mathit{offline}}$ of the partitioning $P_{\mathit{offline}}$ found in the offline phase. If a query $q_i$ takes longer than $-C\cdot r_{\mathit{offline}}/(s_i\cdot f_i)$ we can safely abort it since the corresponding partitioning will definitely result in a lower reward. If we are aware of a partitioning with an even higher reward $r'$, the timeout can further be reduced to $-C\cdot r'/(s_i\cdot f_i).$

\vspace{-1.5ex}\paragraph*{Overall Procedure} Combining all these optimizations, we obtain Algorithm~\ref{alg:online_training} for the online training phase. Note that in case we additionally want to use  timeouts it has to be added in line \ref{alg:line:timeout} of Algorithm~\ref{alg:online_training}.

\subsection{Optimizations for Workload Changes}
\label{sec:dynamic_wl}

In the following, we discuss two enhancements for training the partitioning advisor: (1) using a committee of experts rather than a single agent to further increase the capacity for workloads with many tables and queries, (2) using incremental training to adjust a learned advisor if new queries occur in the workload.

\vspace{-1.5ex}\paragraph*{Committee of Experts} The main goal of our approach is to train a DRL agent just once such that it generalizes over different workload mixes (i.e., different query frequencies). If the workload mix changes, we want to use inference of the trained DRL agent and obtain a new partitioning that works better for the new workload mix.

A \naive{} approach is to train only one DRL agent and sample frequency vectors from a uniform distribution for every episode during training. These frequencies are used both as input of the neural network and in the reward computation during training. The other parts of the training procedure remain unchanged. At inference time, we use the frequency vector as input to our neural network and perform the inference procedure to obtain a partitioning suggested by the agent (as we discuss next in Section \ref{sec:inference}). 

\begin{figure}
    \centering
    \includegraphics[width=0.9\linewidth]{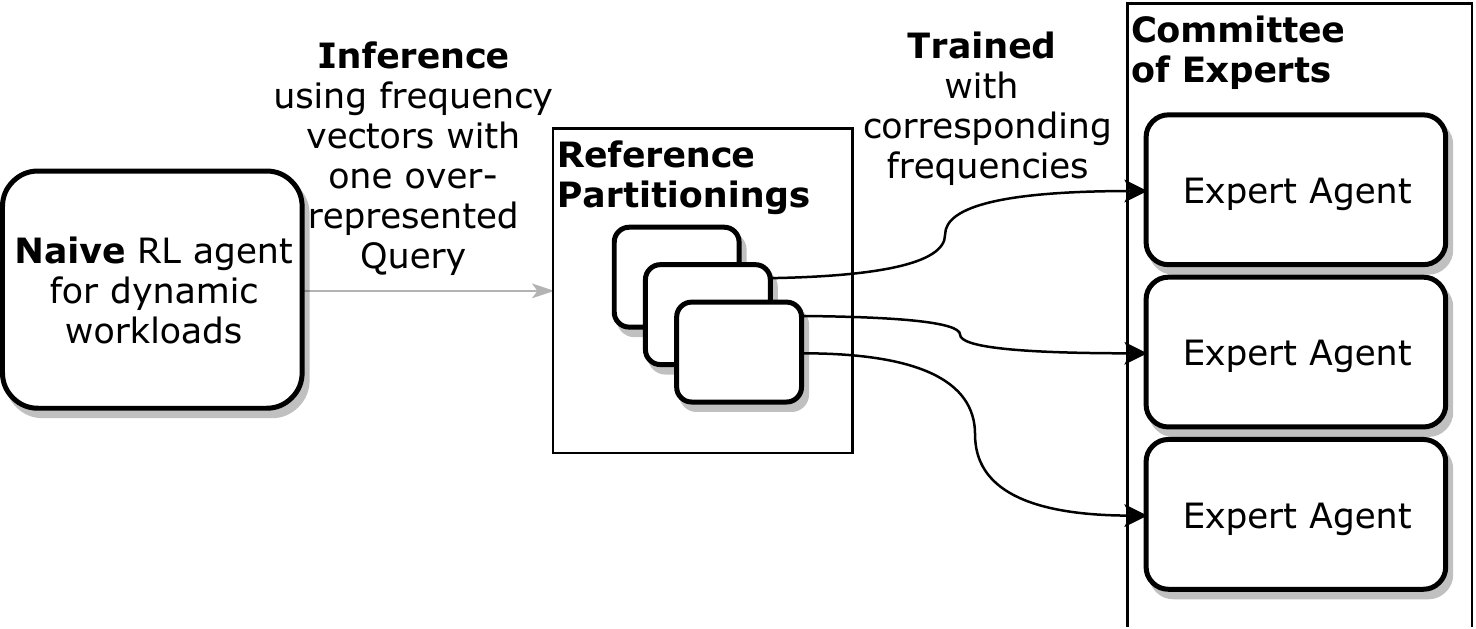}
    \vspace{-0ex}
    \caption{Committee of Experts for Dynamic Workloads.}
    \vspace{-0ex}
    \label{fig:dynamic_wl_experts}
\end{figure}

A more advanced approach enabling more accurate results for a wide variety of workloads (i.e., large query sets) is not to train a single RL agent to suggest partitionings for all possible frequency vectors but to use several expert models for subsets of all possible workload mixes as shown in Figure \ref{fig:dynamic_wl_experts}. The question is how the workload space can be partitioned efficiently into different expert models.
In the following, we explain our approach called \emph{DRL subspace experts}.

The main idea of DRL subspace experts is to first obtain so called reference partitionings $\Tilde{P}_1,\dotsm\Tilde{P}_n$ which represent the set of optimal partitionings.
In order to find these, we use the inference procedure of the \naive{} model (i.e., the RL agent which was trained for the whole workload space) and ask this agent for the optimal partitioning using $m$ frequency vectors where one query $q_i$ is over-represented as shown in the following:

\begin{equation*}
    \begin{pmatrix}
    f_1,\dots,f_{i-1},f_i,f_{i+1},\dots,f_m
    \end{pmatrix}
\end{equation*}

with $f_j = f_{low}$ for $j\in \{0,1,\dots,i-1,i+1,\dots,m\}$ and $f_i=f_{high}$. 
The intuition is that this way we find the optimal partitioning $P_i$ for each individual query.
Since many queries share the same optimal partitioning scheme, the number of distinct partitionings $n$ is much smaller than the number of queries $m$ (i.e., $n << m$).
The distinct partitionings resulting from this step is the set of reference partitionings $\Tilde{P}_1,\dotsm\Tilde{P}_n$ (mentioned before).

For example, for a given workload with $10$ queries we would sample $10$ frequency vectors each representing a workload were one query is over-represented. We then use these to obtain the reference partitionings from a \naive{} model with only one agent. Based on these $10$ frequency vectors, we might end up having just three different reference partitions $\Tilde{P}_1, \Tilde{P}_2$ and $\Tilde{P}_3.$ 

Once we determined the reference partitionings, we can separate the workload space, i.e. the set of different frequency vectors. We say that a frequency vector $(f_1,\dots,f_m)$ belongs to the frequency subspace of one of these reference partitionings $\Tilde{P}_i$ if

\begin{equation*}
\label{eqn_freq_space}
    \Tilde{P}_i=\argmax_{\Tilde{P} \in \{\Tilde{P}_1,\dotsm\Tilde{P}_n\}} -\sum_{j=1}^{m} f_j s_j c_{\mathit{sample}}(\Tilde{P},q_j),
\end{equation*}

i.e. if the reward of the \naive{} RL agent is the maximum among $\Tilde{P}_1,\dotsm\Tilde{P}_n$ for this frequency vector.

Afterwards, we then train one DRL agent for each of these subspaces as depicted in Figure \ref{fig:dynamic_wl_experts}. 
The resulting DRL agents can be considered experts for their frequency subspace. 
For each of the frequency subspaces the training is similar to training the DRL agent for the \naive{} approach. 
The only difference is that the DRL agents are only trained for frequencies of their dedicated subspace. One problem is how to sample more frequency vectors from the same subspace.
To obtain frequency vectors for different subspaces, we sample frequencies uniformly and assign each frequency vector to the DRL agent for the respective reference partitioning $\Tilde{P}_i$.

An important aspect is that the training of these subspace expert models does typically not require any actual execution of queries on the database cluster since we can reuse query runtimes in the Query Runtime Cache of the \naive{} approach. 
When training the subspace expert models, however, we might encounter partitionings that were not seen when training the \naive{} model. 
For these cases, we have no entries in the Query Runtime Cache and the queries need to be actually executed. However, these cases are rare since the \naive{} agent visits all optimal or near-optimal partitionings with high probabilities already. 

\vspace{-1.5ex}\paragraph*{Incremental Training} A final interesting aspect of our online approach is that we can easily support new queries by incremental training.
The main idea is that if new queries are added to a workload, we do not have to train a new model from scratch. 
Instead, we add new inputs representing the query frequencies to the input state of the \naive{} model and retrain it only with frequency vectors that include the new queries. Again, the Query Runtime Cache can be reused and we only have to get actual runtimes for the new queries. Afterwards the \naive{} model can again be used to obtain the new reference partitionings. Only if a new reference partitioning is found, we have to train a new expert agent for that subspace.
Otherwise, it is sufficient to refine the existing subspace experts with the cached query runtimes.

 \section{Model Inference}
\label{sec:inference}

Having trained the DRL agents, we now describe how they can be used to obtain partitionings. In the following, we first assume that only one DRL agent is trained. 
Afterwards, we then explain how the inference works if a committee of experts is used.

\vspace{-1.5ex}\paragraph*{Inference with one DRL agent}
In order to trigger the inference, we assume a frequency vector is given that represents the current workload mix.
When applying the inference procedure, we always start with the same initial state $s_0$ where every table is partitioned by its primary key even if the database is currently partitioned in a different way.
The reason is that starting from the current partitioning might lead to a result where the optimal partitioning is not found since the DRL agent could already be in a local minimum. 
From the initial state $s_0$ as mentioned before, we iteratively choose the action that maximizes the Q-function, i.e. $a_t=\argmax_{a}Q_{\theta}(s_{t+1},a)$. 
This is similar to setting $\varepsilon=0$ or full exploitation and no exploration at all. 
We execute $t_{\max}$ actions and thus obtain a sequence of actions:

\begin{equation*}
    (s_0,a_0,r_0,s_1,\dots,s_{t_{\max}},a_{t_{\max}},r_{t_{\max}}).
\end{equation*}

Afterwards, we do not simply select the partitioning represented by the last state $s_{t_{\max}}$ as optimal partitioning $P^{\ast}$, since the DRL agent tends to oscillate around the optimal partitioning $P^{\ast}$ (i.e, the partitioning with the highest reward is not necessarily represented by the last state). 
Instead, we identify the state $s_t$ in the sequence above with a maximum reward and return the corresponding partitioning $P^{\ast}.$

\vspace{-1.5ex}\paragraph*{Inference with a committee of DRL agents}

If we want to obtain a new partitioning when a committee of experts was trained, we first determine which subspace $\Tilde{P}_i$ of the frequency space the vector belongs to:

\begin{equation*}
    \Tilde{P}_i=\argmax_{\Tilde{P} \in \{\Tilde{P}_1,\dotsm\Tilde{P}_n\}} -\sum_{j=1}^{m} f_j s_j c_{\mathit{sample}}(\Tilde{P},q_j).
\end{equation*}

The DRL agent is selected by choosing the DRL agent for the reference partitioning with the lowest estimated runtime (which is the same procedure we use when training the expert models).
Afterwards, we use the inference procedure discussed before with the corresponding expert model for~$\Tilde{P}_i.$ 

\section{Experimental Evaluation}
\label{sec:experiments}

In the following, we evaluate the benefits of using learned partitioning advisors for databases with schemas of varying complexity. 
In the first experiment (Section \ref{sec:experiments:exp1}), we validate that DRL agents that are trained purely offline find partitionings outperforming typical heuristics and are competitive with those found by state-of-the-art partitioning advisors such as \cite{nehme2011automated}. 
Furthermore, if additionally trained online, the DRL agent clearly outperforms state-of-the-art systems and finds non-obvious partitionings with superior runtime as we demonstrate in our second experiment (Section \ref{sec:experiments:exp2}). Another benefit of our approach is the flexibility w.r.t. changes in the workload (Section \ref{sec:experiments:exp3}). Hence, in the third experiment we first show that the committee of experts can suggest partitionings that improve over the \naive{} model for changing workloads. Furthermore, we examine the additional training time required if new queries are added to a workload. Finally, in the last experiment (Section \ref{sec:experiments:exp4}) we show that our agent can also adapt to changes in the deployment (i.e., if hardware characteristics change) which is not trivial with existing approaches.

\begin{table}
\begin{scriptsize}
\centering
\begin{tabular}{ll}\hline
\textbf{Parameter} & \textbf{Value} \\\hline
Learning Rate & $5\cdot 10^{-4}$ \\
$\tau$ (Target network update) & $10^{-3}$ \\
Optimizer  & Adam \\
Experience Replay Buffer Size & 10000 \\
Batch Size for Experience Replay & 32 \\
Epsilon Decay & 0.997 \\
$t_{\max}$ (Max Stepsize) & 100 \\
Episodes & 600/1200 \\
Network Layout & (128,64) \\
$\gamma$ (Reward Discount) & 0.99 \\\hline
\end{tabular}
\caption{Hyperparameters used for DRL training.}
\label{table:hyperparameters}
\vspace{-2.5ex}
\end{scriptsize}
\end{table}

\subsection{Workloads, Setup and Baselines}

For the experiments, we used different databases and workloads that we explain in the following. Moreover, we also discuss the learning setup that we used for training the partitioning advisors as well as the baselines we compared to.

\vspace{-1.5ex}\paragraph*{Data and Workloads}
 
We evaluated the partitioning advisor on three different database schemas and workloads varying in complexity:
(1) As the most simple case, we used the Star Schema Benchmark (SSB) and its workload \cite{o2007star}. SSB is based on TPC-H and re-organizes the database into a pure star schema with $5$ tables ($1$ fact and $4$ dimension tables) and $13$ queries.
(2) The second database and workload we used was TPC-DS \cite{tpcds}. TPC-DS comes with a much more complex schema of $24$ tables ($7$ fact and $17$ dimensions tables) and $99$ queries (including complex nested queries). 
(3) In cloud data warehouses such as Amazon Redshift, customers are not required to use a star schema but can design an arbitrary schema for their database. 
To test how well our learned advisor can cope with more complex schemata which are not based on a star-schema, we additionally used the TPC-CH benchmark \cite{funke2011benchmarking}, which is the combination of the schema of the TPC-C benchmark with analytical queries of the TPC-H schema (adopted for the TPC-C schema). 
Originally, the TPC-CH benchmark combines analytical queries and transactions in a mixed workload. 
For the purpose of this paper, we only used the analytical queries to represent the workload in our evaluation. Furthermore, in the standard version of TPC-CH all tables can be co-partitioned by the \texttt{warehouse-id}. While our DRL agents also propose this solution when using the original TPC-CH schema, we do not think that such a trivial solution is realistic for many real-world schemata. Hence, we further added complexity and decided to restrict possible partitionings such that tables cannot be partitioned by \texttt{warehouse-id} only.

\vspace{-1.5ex}\paragraph*{Learning Setup}

The partitionings for different analytical schemas were evaluated on two database systems.
To show that our learned approach is in general applicable to both disk-based and memory-based distributed databases, we used Postgres-XL (a popular open-source distributed disk-based database) \cite{pgxl} and System-X (a commercial distributed in-memory database). 
For running the databases in a distributed setup, we used CloudLab \cite{cloudlab}, a scientific infrastructure for research on cloud computing. 
For our experiments, we provisioned clusters of different sizes ranging from $4$ to $6$ nodes. 
Each node was configured to use $128$GB of DDR4 main memory, two Intel Xeon Silver 4114 10-core CPUs and a $10$Gbps interconnect. 

For learning the partitioning advisor, we used neural networks implemented in Keras.
In particular, the neural network to approximate the Q-functions used $2$ hidden layers with $128$ and $64$ neurons, respectively.
We used the standard ReLU activation function in every layer and a linear function for the output (to represent the Q-value) which is a common combination for DRL. 
Furthermore, we used Adam as optimizer which adapts the learning rate for every parameter. 
An overview of all hyperparameters which we found to work best for training can be seen in Table \ref{table:hyperparameters}.
The only hyperparameter we changed for the different databases was the amount of episodes we used to train the model. 
Since SSB has a significantly lower amount of tables and queries we only trained the DRL agents for $600$ episodes instead of $1200$ episodes for TPC-DS and TPC-CH.

\vspace{-1.5ex}\paragraph*{Baselines} We first compared the partitionings found by our approaches to heuristics that are typically used by a database administrator \cite{localityPartitioning}. 
For both simple and more complex star schemata (SSB and TPC-DS) this means that usually fact tables are co-partitioned with either the most frequently joined dimension table (Heuristic~1) or the largest dimension table (Heuristic~2). For the more complex schema TPC-CH, we either \naively{} replicated small tables and partitioned larger tables by primary key (Heuristic~1) or greedily co-partitioned the largest pairs of tables while still replicating smaller tables (Heuristic~2).

We also compared our DRL-based approach to an automatic partitioning approach that mimics \cite{nehme2011automated} and uses the same cost-model to estimate the runtime of queries to find a partitioning scheme given a schema and a workload.
As a result we saw that when using only the offline-training (Exp. 1), our DRL-based approach finds results competitive with \cite{nehme2011automated} which is what we expected.
However, when training the DRL agent additionally with the online phase (Exp. 2 - 4), our advisor finds non-obvious solutions since it can better adapt to the actual execution cost and different hardware setups.

\begin{figure}
  \centering
  \begin{subfigure}[b]{0.48\linewidth}
   \centering
   \includegraphics[width=1\linewidth]{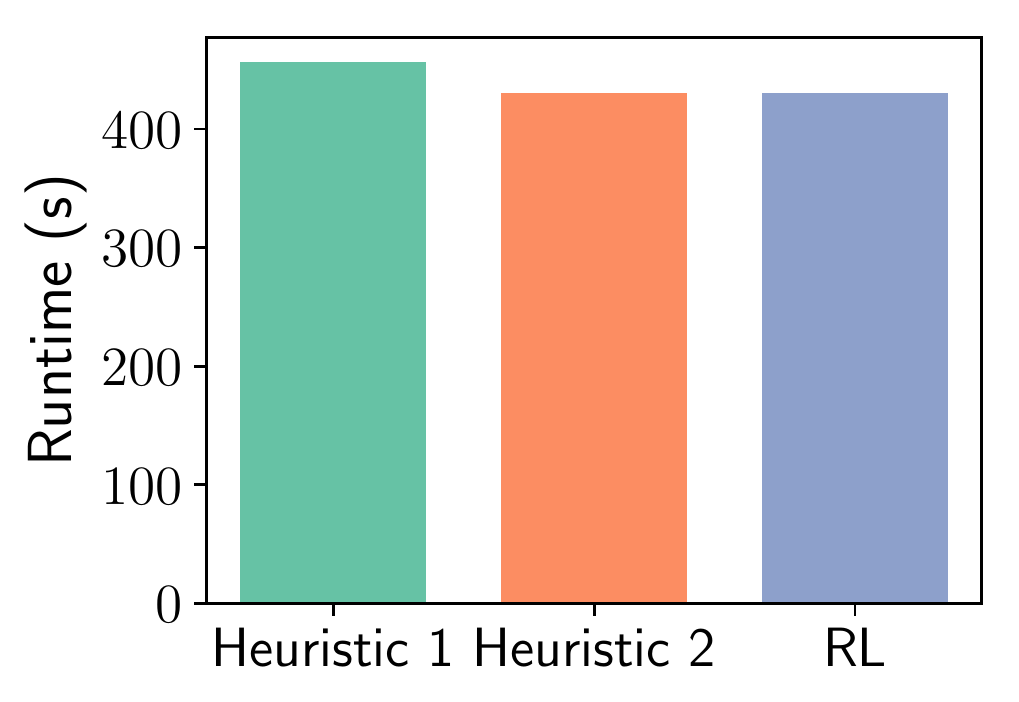}
   \caption{SSB (Postgres-XL)}
  \end{subfigure}
  \begin{subfigure}[b]{0.48\linewidth}
   \centering
   \includegraphics[width=1\linewidth]{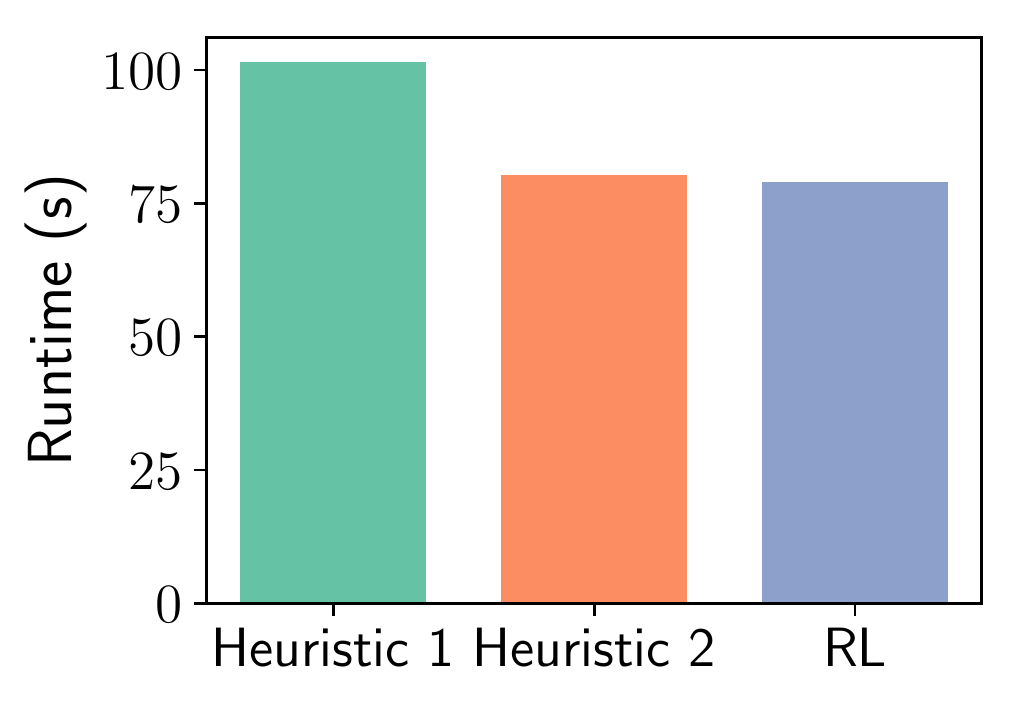}
   \caption{SSB (System-X)}
  \end{subfigure}
  
  \begin{subfigure}[b]{0.48\linewidth}
   \centering
   \includegraphics[width=1\linewidth]{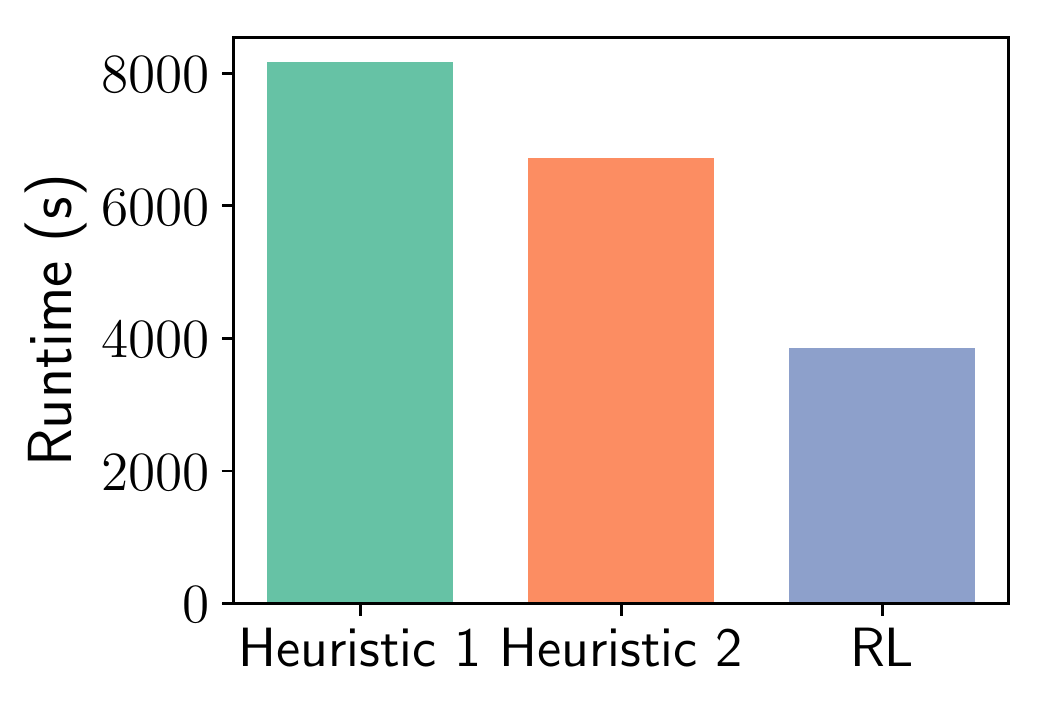}
   \caption{TPC-DS (Postgres-XL)}
  \end{subfigure}
  \begin{subfigure}[b]{0.48\linewidth}
   \centering
   \includegraphics[width=1\linewidth]{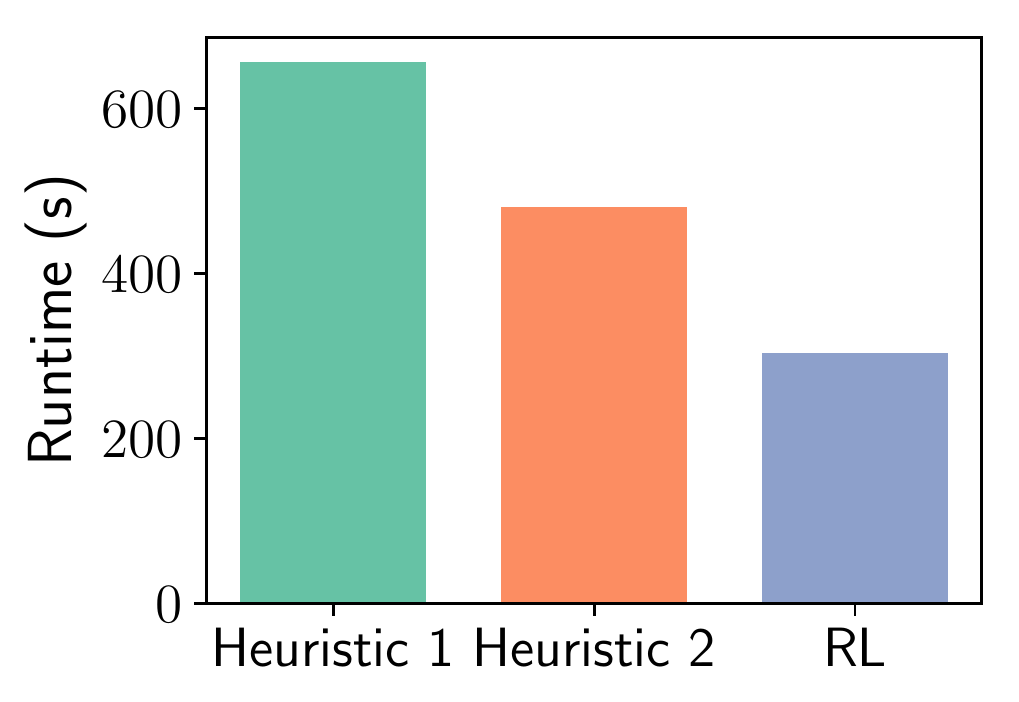}
   \caption{TPC-DS (System-X)}
  \end{subfigure}
  
  \begin{subfigure}[b]{0.48\linewidth}
   \centering
   \includegraphics[width=1\linewidth]{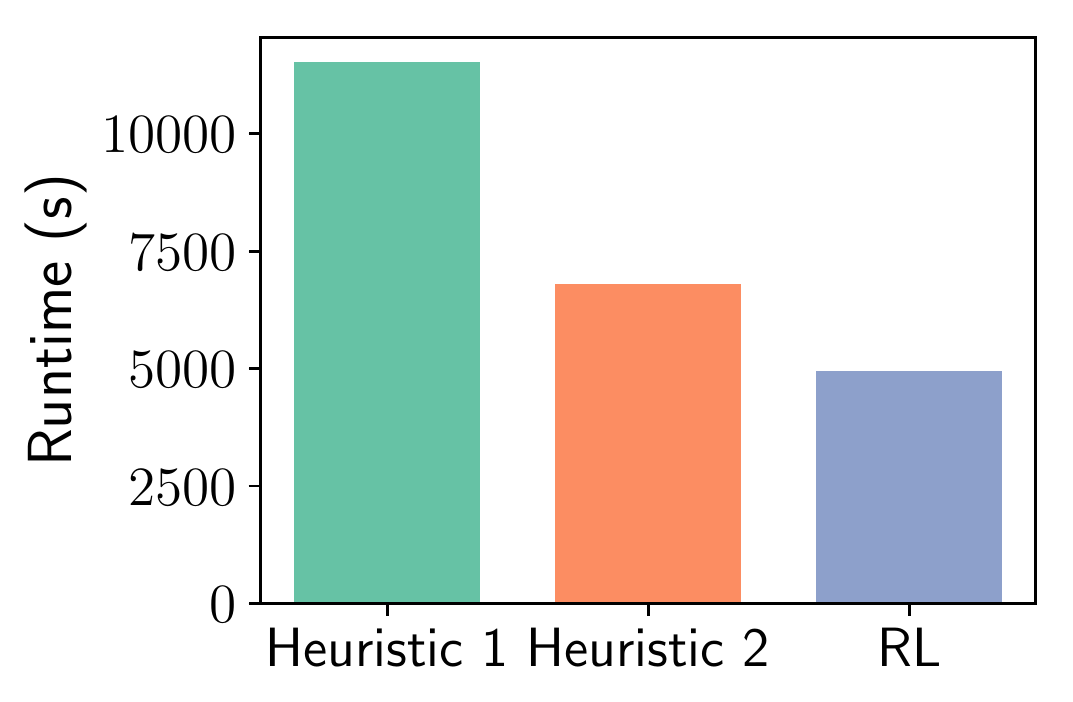}
   \caption{TPC-CH (Postgres-XL)}
  \end{subfigure}
  \begin{subfigure}[b]{0.48\linewidth}
   \centering
   \includegraphics[width=1\linewidth]{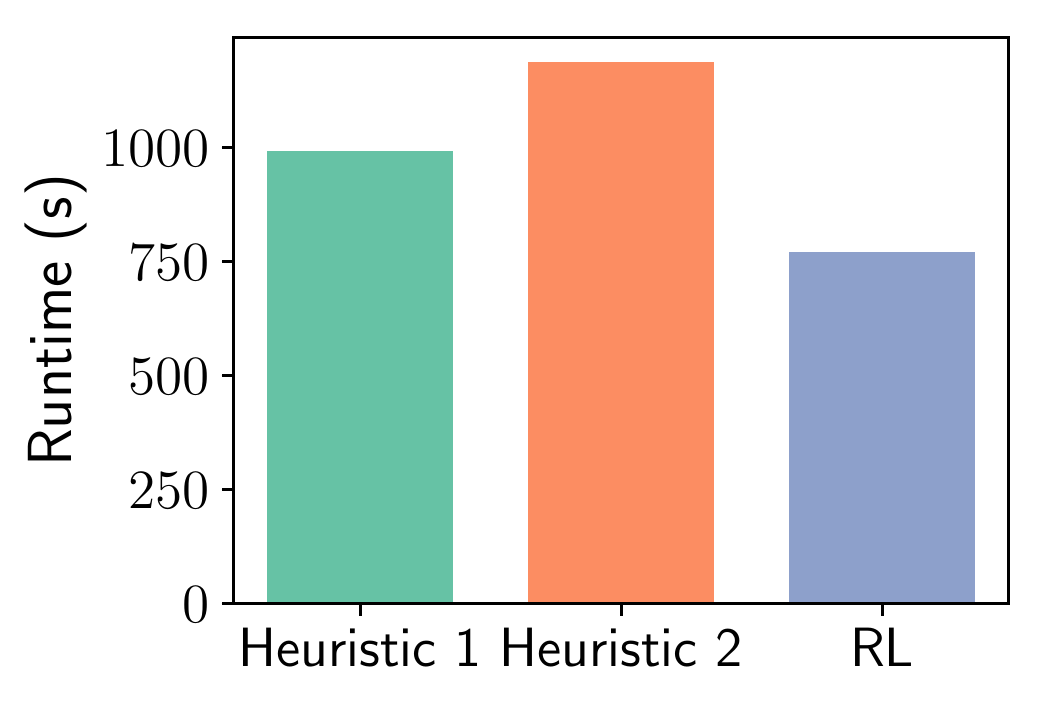}
   \caption{TPC-CH (System-X)}
  \end{subfigure}
\caption{Runtimes with partitionings proposed by DRL agent (trained offline) for Postgres-XL and System-X.}
\label{fig:experiments_offline}
\end{figure}

\subsection{Exp. 1: Offline Training}
\label{sec:experiments:exp1}

For each database mentioned before in the setup, we trained a dedicated DRL agent with offline training, i.e. using a cost model to estimate the runtime. 
For all databases and workloads (SSB, TPC-DS, and TPCH-CH), we used the scaling factor SF=100 and for the workloads we assumed that all queries occur equally frequently. 
As a result we report the total runtime of all queries in the entire workload for the partitionings suggested by our advisor. 
For Postgres-XL, only $60$ of the $99$ queries could be executed due to restrictions in which queries it can support.
The runtimes of our DRL-based approaches for Postgres-XL and System-X compared to the baselines are given in Figure \ref{fig:experiments_offline}.

\vspace{-1.5ex}\paragraph*{Results for SSB} For the SSB benchmark, the two heuristics co-partition the fact table with either the most frequently joined dimension table (\texttt{Date}) or the largest dimension table (\texttt{Customer}). Our learned advisor also suggests to co-partition the fact table with the largest dimension table for Postgres-XL (same as Heuristic~2). For System-X our learned advisor additionally suggests to partition the \texttt{Part} dimension table by its primary key leading to a minimal runtime improvement.

\vspace{-1.5ex}\paragraph*{Results for TPC-DS} For TPC-DS, which is a more complex schema composed of several fact tables with shared dimensions, the DRL agents finds superior solutions that are non-obvious. 
Here, the improvements are more significant reducing the runtime over Heuristic~1 by approximately 50\%. 
For both Postgres-XL and System-X, the DRL agents propose to co-partition the fact tables with a medium-sized dimension table, i.e. \texttt{Item}. 
This has the advantage that local joins are possible if two fact tables are joined, e.g. the fact tables for \texttt{StoreSales} and \texttt{StoreReturns}. 
Moreover, for System-X the \texttt{Customer} table is co-partitioned with the \texttt{CustomerAddress} table allowing local joins.

\begin{figure}
\centering
\includegraphics[width=1\linewidth]{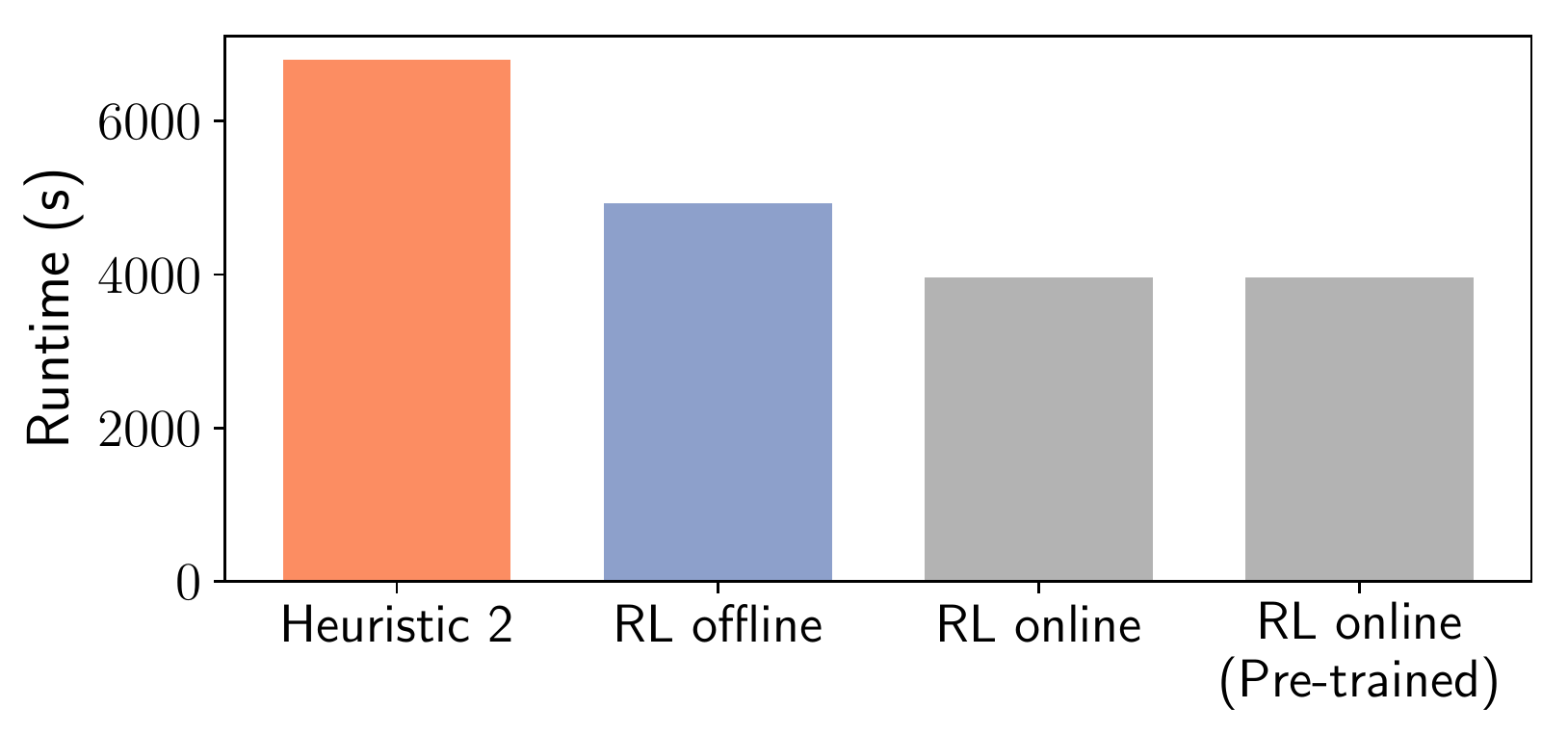}
\vspace{-1.5ex}
\caption{Runtimes with different partitionings proposed by the DRL agent (trained online) compared to baselines.}
\label{fig:experiments_pgxl_online}
\end{figure}

\vspace{-1.5ex}\paragraph*{Results for TPC-CH} As discussed before, TPC-CH uses a significantly more complex schema than SSB and TPC-DS since it is not similar to a star schema. 
While Heuristic~2 has better runtimes than Heuristic~1 on Postgres-XL, Heuristic~1 outperforms Heuristic~2 on System-X. This counter-intuitive result is due to the fact that partitioning a table by \texttt{district-id} resulted in skewed partition sizes in System-X, which significantly reduced the runtime of Heuristic~2.
Compared to the two heuristics, the DRL-agent proposes improved partitionings. For Postgres-XL it proposes to co-partition the \texttt{Customer}, \texttt{Order}, \texttt{NewOrder} and additionally the \texttt{Orderline} table by \texttt{district-id} but to replicate the \texttt{Stock} table. This avoids that \texttt{Orderline} has to be shuffled over the network for a join. 
For System-X, the DRL agent additionally partitioned the \texttt{Stock} table but also used a compound key combining \texttt{warehouse-id} and \texttt{district-id} to mitigate the skew (which was reflected in the cost model).

\subsection{Exp. 2: Online Training}
\label{sec:experiments:exp2}

In this experiment we show that DRL agents trained online are superior over purely offline-trained agents.  
For showing this, we focus on the most complex schema, i.e. TPC-CH, to analyze the benefits of the additional online phase that can leverage actual costs instead of cost estimates to refine the partitioning advisor.
For the online training we started with a lower $\varepsilon$-value which is a common technique if less exploration is needed. 
To demonstrate that offline training reduces the training time for the online model, we also compare the training time to the time needed to train a model online without any offline pre-training. 

\begin{figure}
\centering
\includegraphics[width=1\linewidth]{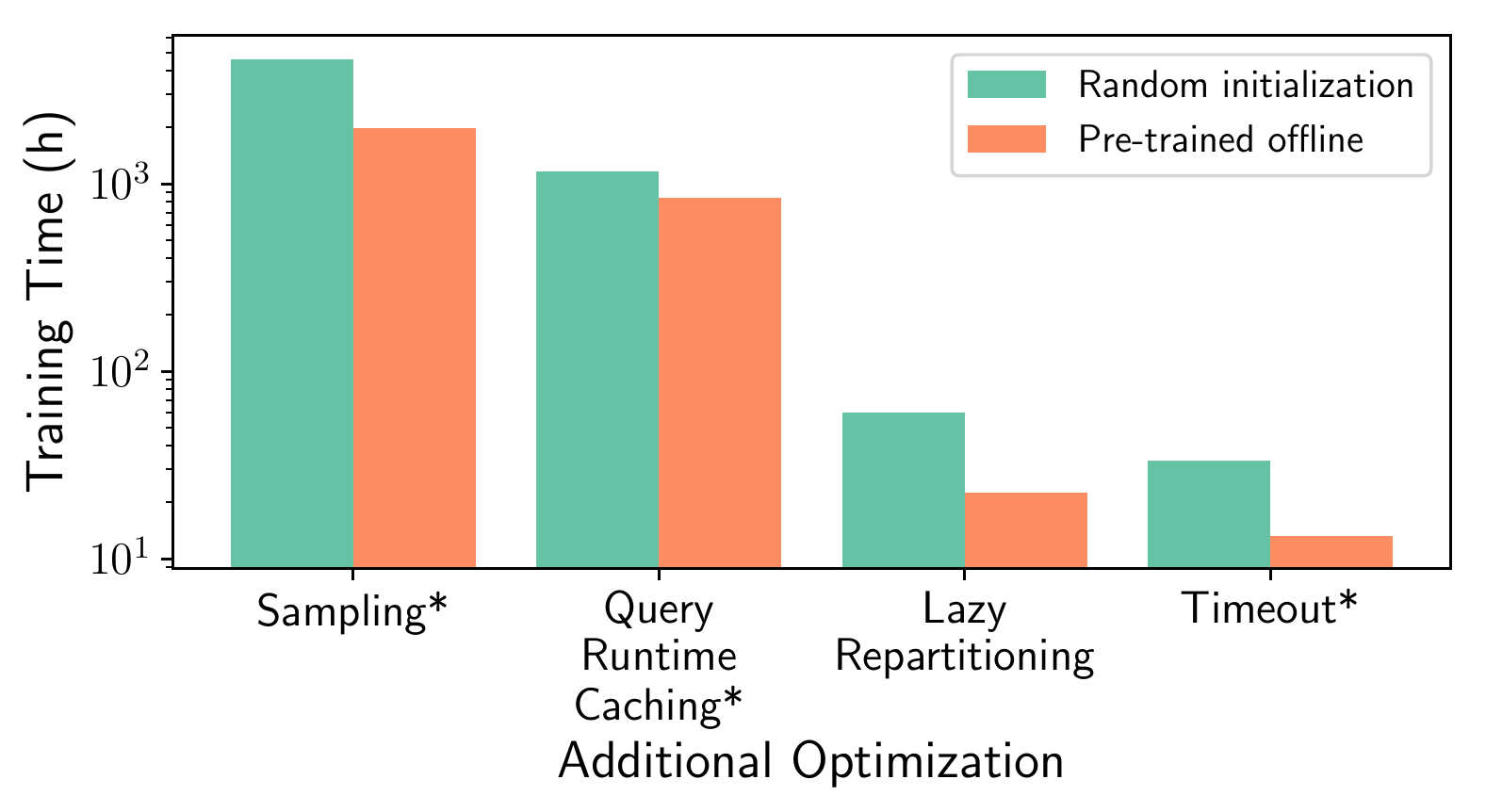}
\caption{Training Time for Online Phase.}
\label{fig:optimizations_online}
\end{figure}

The results of using the suggested partitioning on the full TPC-CH database are shown in Figure~\ref{fig:experiments_pgxl_online}. 
Both DRL agents (with and without offline pre-training) suggested the same partitioning and hence result in the same runtime. 
However, when comparing the online-trained agent with the agent that was only trained offline, we could observe that the runtime of the workload was improved. 
Different from the offline-trained agent, after applying the online phase the DRL agent suggests a new partitioning where the \texttt{NewOrder}, \texttt{Order} and \texttt{Orderline} table are co-partitioned by \texttt{Order-Id} and the \texttt{Customer} table is replicated in addition. Interestingly, this partitioning has higher costs according to the cost model. 
However, the online phase is not affected by the inaccuracy of the cost model and was thus able to improve over the offline-trained agent.

\begin{table}
\begin{scriptsize}
\centering
\begin{tabularx}{0.45\textwidth}{Xll}\hline
\textbf{Additional Optimization} & \textbf{Runtime} & \textbf{Improvement} \\\hline
Sampling & 1972.6h & - \\
Query Runtime Cache & 843.6h & 133.8\% \\
Lazy Repartitioning & 22.5h & 3655.6\% \\
Timeouts & 13.3h & 69\% \\
\hline
\end{tabularx}
\caption{Training Time Reduction (Pre-trained offline).}
\label{table:optimizations_online}
\end{scriptsize}
\end{table}

If executed \naively{}, the online phase is time-consuming. As mentioned above, we ran the online training procedure twice: once for a randomly initialized model and once for the pre-trained model. Moreover, we want to examine the effect of different optimizations. 
The training time of the online phase can be seen in Table~\ref{table:optimizations_online} and Figure~\ref{fig:optimizations_online}.
For this experiment, we were only running the training with all optimizations (except timeouts) activated. We then derived from the logs how long the total training would have taken without the optimizations. As we can see every optimization significantly reduces the runtime and the largest improvement can be obtained with Lazy Repartitioning.

The online-phase with all optimizations and for a model that was pre-trained offline took approximately 13.3 hours. We believe that a training time of several hours is acceptable since the model only has to be trained once for different workload mixes and can afterwards be used as a partitioning advisor if the workload changes (as we show in the next experiment). 
Moreover, especially in cloud setups, we can easily clone the instances. Hence, setting up a similar cluster to retrain the agent for several hours to obtain a refined model should be feasible considering that customers usually have one cluster provisioned all the time to do analytics.

\subsection{Exp. 3: Adaptivity to Workload}
\label{sec:experiments:exp3}

The following experiments validate the adaptivity of a DRL agent to changing workloads.
We first demonstrate that our approach finds optimal partitionings without additional training if only the query mix changes. Moreover, in a second experiment we examine the additional training time required if completely new queries are added to the workload.

\begin{figure}
\includegraphics[width=1\columnwidth]{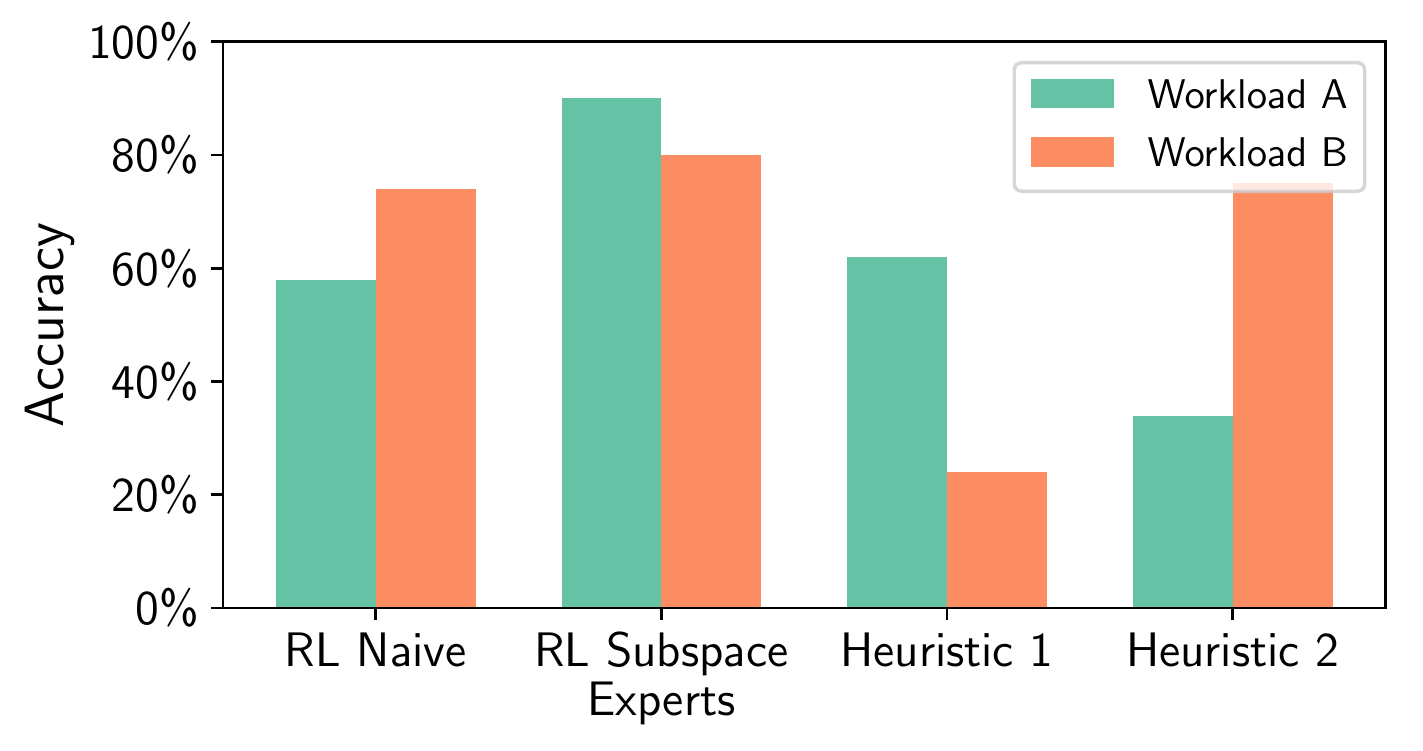}
\caption{Optimal Partitioning found by Different Approaches (higher is better).}
\label{fig:workload_adaptivity}
\end{figure}

\vspace{-1.5ex}\paragraph*{Exp. 3a: Changing Workload Mix} In this experiment we show that our learned advisor finds optimal partitionings for different query mixes. To this end, we trained an DRL agent with the \naive{} approach for different workload frequencies for the TPC-CH schema. Moreover, we additionally trained a committee of experts for the subspace experts approach as described in Section \ref{sec:dynamic_wl}. 

For the \naive{} model and the committee of experts, we both used an online training phase on Postgres-XL. Note that we can apply all optimizations for the online training as well. Moreover, we can reuse the Query Runtime Cache of the previous experiments if we train multiple experts. In order to be fair to the \naive{} model, we decided to increase its capacity and thus used 256 and 128 neurons in the hidden layers and extend the training time to $2000$ episodes. 

After training both approaches, we evaluated how often these approaches find the optimal partitioning for two different cluster workload mixes. Each cluster is a set of different frequency vectors (i.e., workload mixes): for cluster A the frequencies were sampled uniformly and for cluster B queries joining the \texttt{Stock} and the \texttt{Item} tables are more likely to occur.
If the partitioning found by either approach is best for the respective cluster, we say that the approach has found the optimal partitioning for this workload mix. 

We compare the \naive{} approach and the subspace experts approach with two heuristics. Heuristic~1 always chooses the optimal partitioning found after online training in the previous Section. Heuristic~2 always chooses a partitioning where the \texttt{Stock} and \texttt{Item} tables are co-partitioned. 
The results are given in Figure \ref{fig:workload_adaptivity}. As we can see, the accuracy can significantly be improved when using the subspace experts approach outperforming all other approaches for every workload. We conclude that is beneficial to divide the problem of finding an optimal partitioning for a given workload into subproblems which are then solved by the dedicated expert model.

\vspace{-1.5ex}\paragraph*{Exp. 3b: New Queries} In this experiment, we evaluate the additional training overhead if new queries are introduced. In particular, it shows that additional training is much cheaper than training the agent from scratch if new queries occur.

We again trained a committee of experts for TPC-CH on top of Postgres-XL as the underlying database.
However in contrast to the previous experiment, we first randomly removed a fraction of the queries of the TPC-CH benchmark. We then retrained the advisor for the additional queries and calculated, with the help of already measured runtimes, how long such an additional training takes on average if part of the workload is not known initially.

\begin{figure}
\centering
\includegraphics[width=1\linewidth]{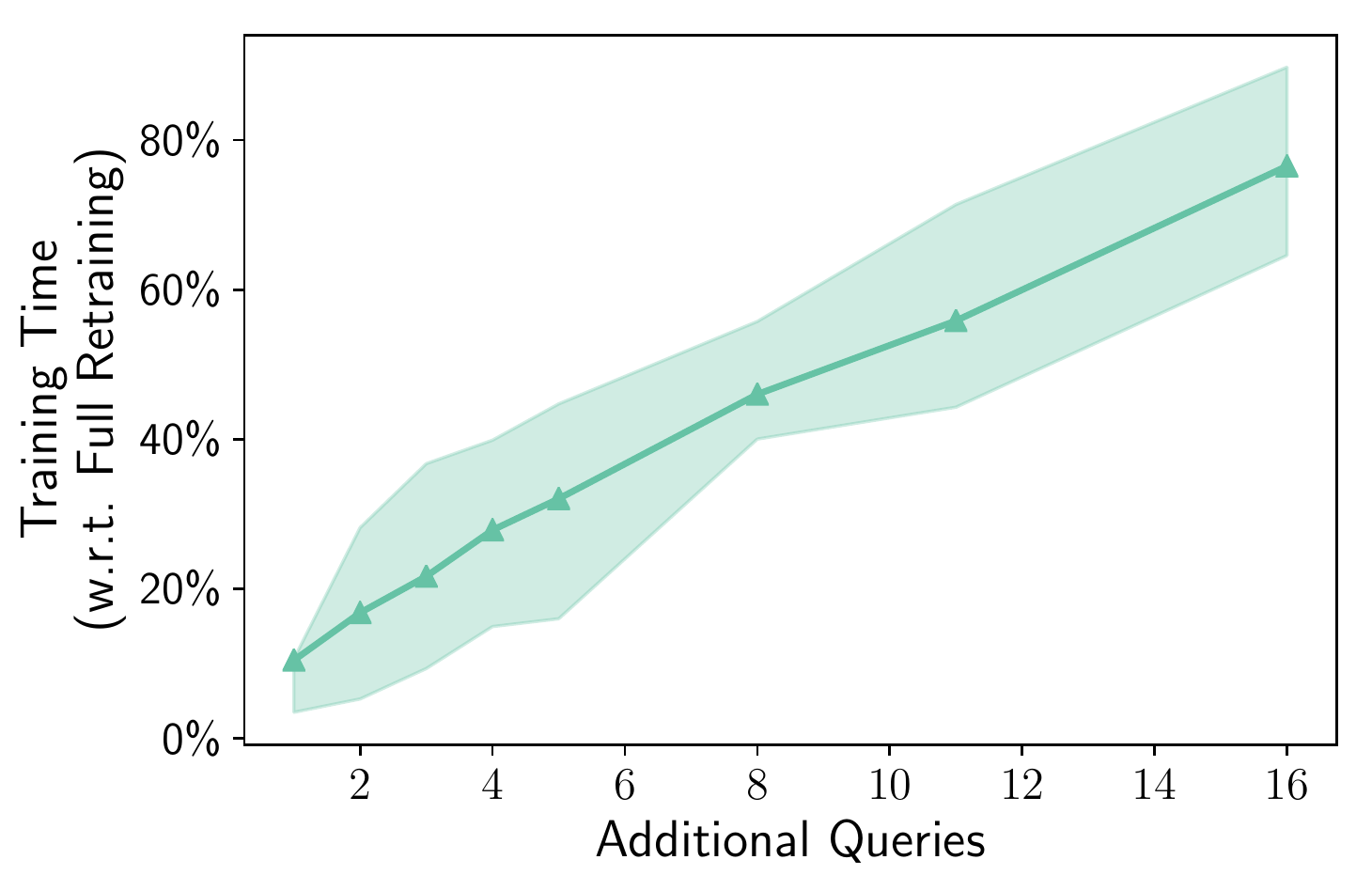}
\caption{Training Time of Additional Training (relative to Full Retraining) with 25\% and 75\% Quantiles.}
\label{fig:delta_train}
\end{figure}

Figure~\ref{fig:delta_train} shows the time for incremental training relative to the time required to train an DRL agent from scratch, depending on how many additional TPC-CH queries were added after the initial training.
As we can see, the overhead of incremental training is much lower than training a partitioning advisor from scratch. This is because, similar to exploiting a bootstrapped DRL agent using the offline phase, we can start with a lower $\varepsilon$-value in the incremental training of the new \naive{} model resulting in fewer explorations. In addition, incremental training can also make use of the Query Runtime Cache, which keeps actual query execution to a minimum as many queries are already known.

Another interesting observation is that in most cases, the incremental training did not result in extending the committee of experts. Only in one case, which added two particular new queries that join two fact tables, an expert for a new partitioning was added to the learned advisor. If we use inference for a query mix where these two new queries occur more frequently than the old queries the new partitioning is 25\% faster on the full dataset compared to the partitioning scheme that was suggested without retraining.

\subsection{Exp. 4: Adaptivity to Deployment}\label{sec:experiments:exp4}

Another advantage of using an DRL agent as partitioning advisor is that it can adapt the partitioning for different deployments which is an important scenario for  cloud providers that allow customers to migrate their cluster to a new set of virtual machines with different characteristics. 
For showing the adaptivity of our learned advisor, we created a simple microbenchmark to empirically validate this. 
It consists of three relations \texttt{A}, \texttt{B} and \texttt{C} where \texttt{A} is a fact table and \texttt{B} and \texttt{C} are dimension tables. The relation sizes are given in Table~\ref{table:relation_sizes_microbenchmark} and are inspired by the relation sizes of the \texttt{Lineorder}, \texttt{Order} and \texttt{Partsupp} table of the TPC-H benchmark. The workload consists of just two queries joining the fact table~\texttt{A} with one of the dimension tables~\texttt{B} or \texttt{C} with selectivities between 2\% and 5\%. 

In the optimal partitioning, table \texttt{A} and \texttt{C} have to be co-partitioned because \texttt{C} is significantly larger than \texttt{B}. Depending on the network bandwidth, however, it might be optimal to either partition or replicate table~\texttt{B}. For example, for a high-bandwidth network it might be beneficial to partition \texttt{B}, say, on its primary key. When joined with table~\texttt{A} the scan of table~\texttt{B} can be distributed among all cluster nodes (if the table is partitioned) and the remaining tuples have to be shuffled over the network. If table~\texttt{B}, however, is replicated we do not have to send tuples over the network for the join but the scan is also not distributed across nodes. Hence, the question whether or not partitioning is beneficial depends on the speed of the network compared to the scan speed of the table. As network costs are more significant if one does not need costly disk accesses we decided to use System-X for the evaluation which is an in-memory database.

\begin{figure}
\centering
\includegraphics[width=1\linewidth]{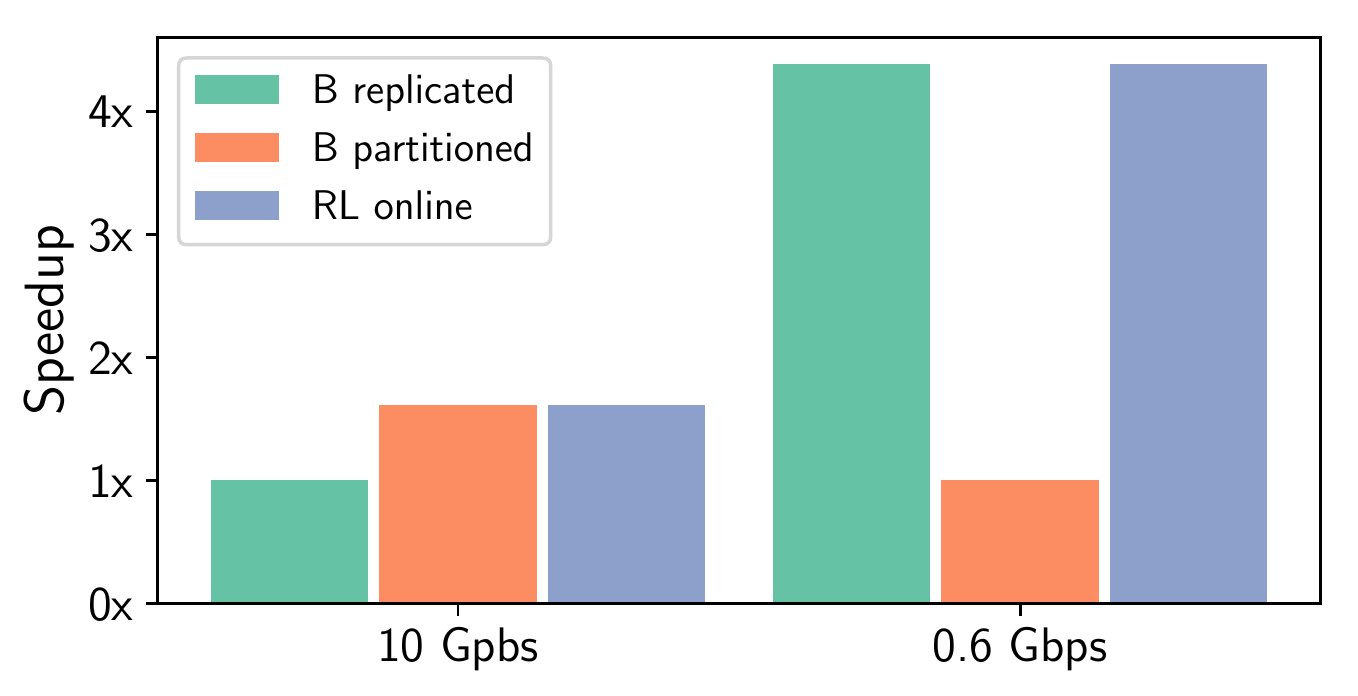}
\caption{Adaptivity to Network Speeds (higher is better).}
\label{fig:hardware_deployments}
\end{figure}

To show the effect, we used two different hardware deployments for System-X. One time, we used the usual 10 Gbps interconnect, one time we only used 0.6 Gbps interconnects. 
This is also the bandwidth offered for the basic deployment of Amazon Redshift. 
We trained one DRL agent on the full dataset (approx. 100 GB) for the two hardware deployments. 
In Figure~\ref{fig:hardware_deployments} the effects of partitioning or replicating table~\texttt{B} can be seen for both the slow and the fast network.
In the figure, we use the slowest approach of both as reference and show the speed-up of the others (i.e., higher is better).
As we can see, for the slow network it is optimal to replicate table~\texttt{B}, while for the fast network it is better to partition it. 
In both cases the DRL agent (after retraining the model on the hardware setup using our online phase) suggests the optimal solution.
 
\begin{table}
\begin{scriptsize}
\centering
\begin{tabular}{ll}\hline
\textbf{Relation} & \textbf{Tuples in Millions}\\\hline
A & $\approx134.2$ \\
B & $\approx6.7$ \\
C & $\approx33.6$ \\
\hline
\end{tabular}
\caption{Relation sizes of Microbenchmark.}
\label{table:relation_sizes_microbenchmark}
\end{scriptsize}
\end{table}

 \section{Related Work}
\label{sec:related_work}

Finding optimal partitionings is an active field of research. In the following, we discuss related work in different categories:

\paragraph*{Partitioning for OLTP and OLAP} Many approaches focus on transactional workloads \cite{pavlo2012skew,Curino2010SchismAW,Fetai2015WorkloaddrivenAD,Chen2015OnlineDP, quamar2013sword}. In general, these approaches try to partition the data such that distributed transactions across nodes occur less frequently. For example, SCHISM \cite{Curino2010SchismAW} defines a graph consisting of tuples as nodes and transactions as edges and uses a min-cut to partition the tuples. Pavlo et al. \cite{pavlo2012skew} developed an alternative approach that is also capable of stored procedure routing and replicated secondary indexes. Fetai et al. focus especially on cloud environments \cite{Fetai2015WorkloaddrivenAD}.

For OLAP-workloads Eadon et al. \cite{eadon2008supporting} proposed \texttt{REF}-parti\-tioning, i.e. to co-partition chains of tables linked via foreign key relationships. Since this technique can be exploited if a system supports hash-partitioning by any attribute most partitioning advisors and also our technique indirectly make use of \texttt{REF}-partitioning. 

Zamanian et al. \cite{localityPartitioning} extend this approach such that even more locality can be obtained but at the cost of higher replication. For this, the database has to support predicate-based reference partitioning. In contrast, \cite{lu2017adaptdb} iteratively improves the partitioning and relies on hyper-partitioning and hyperjoins as database features. However, these features are currently not supported by Postgres-XL or System-X and hence could not be evaluated as a baseline.

\vspace{-1.5ex}\paragraph*{Automated Database Design} Automatic design advisors are an active area of research \cite{nehme2011automated,DBLP:conf/sigmod/RablJ17,DBLP:conf/vldb/ZilioRLLSGF04,rao2002automating}. 
However, many of these approaches \cite{DBLP:conf/vldb/ZilioRLLSGF04,rao2002automating} focus only on single-node systems while only a few devised advisors for distributed databases are specialized on partitioning design \cite{nehme2011automated,DBLP:conf/sigmod/RablJ17}.
These approaches, however, rely only on cost models and are thus comparable to the offline phase presented in this paper. Different from these approaches our approach implements a dedicated online phase that is able to cope with inaccuracies of the cost model.

Another approach \cite{DBLP:conf/sigmod/RablJ17} optimizes both analytical and transactional workloads by partially allocating already partitioned tables in an optimal manner to minimize runtime or maximize throughput.
Different from this approach, which is only focusing on the allocation, in this paper we provide a new solution to find a partitioning scheme which is an orthogonal problem to data allocation.
Furthermore, the paper relies on an allocation heuristic which cannot take the actual execution cost into account.

Recently, many approaches suggest to use machine learning to automate database administration and tuning \cite{kraska2019sagedb,pavlo2017self} and improve internal database components like join ordering \cite{krishnan2018learning} or cardinality estimation \cite{kipflearned}. In particular, DRL \cite{silver2016mastering,mnih2015human} was often used to tackle data management problems. For example Li et al. \cite{li2018model} focused on the scheduling problem for distributed stream data processing systems or Durand et al. \cite{durand2018gridformation} optimized the physical table layout.
Different from those papers, we focus on data partitioning in distributed databases which was not yet considered.

 \section{Conclusion and Future Work}
\label{sec:conclusion}

In this paper, we introduced a new approach for learning a partitioning advisor based on DRL. 
The main idea is that a DRL agent learns its decisions based on experience by monitoring the rewards for different workloads and partitioning schemes.
In the evaluation, we showed that our approach is not only able to find partitionings that outperform existing approaches for automated partitioning design but that it can also easily adjust to different deployments. 

In the future, we plan to combine our approach with systems that predict future workloads to pro-actively re-partition the database as well as to decide whether the costs for repartitioning pay off in the long run. 
Another interesting avenue of future work is to use our approach for transactional workloads or combined systems for analytical and transactional workloads. 
\balance{}
\bibliographystyle{abbrv}
\bibliography{bib}

\end{document}